\documentclass[a4paper,fleqn,usenatbib]{mn2e}

\usepackage{txfonts}
\usepackage[T1]{fontenc}
\usepackage{times}
\usepackage{BibDef}
\usepackage{hyperref}
\usepackage{amssymb}
\usepackage{float}
\usepackage{color}
\usepackage{subfigure}
\usepackage{graphicx}
\usepackage{soul}

\def\be{\begin{equation}}
\def\ee{\end{equation}} 
\def\ltsima{$\; \buildrel < \over \sim \;$}
\def\lsim{\lower.5ex\hbox{\ltsima}}
\def\gtsima{$\; \buildrel > \over \sim \;$}
\def\gsim{\lower.5ex\hbox{\gtsima}}

\def\vg{$V_{\rm G}~$}
\def\vgs{V_{\rm G}}
\def\kms{km s$^{-1}$}
\def\rs{$r_{\rm s}~$}
\def\rss{r_{\rm s}}
\def\Mhs{M_{\rm h}}

\title[Constraints from HVSs]{Joint constraints on the Galactic dark matter halo and GC from hypervelocity stars}

\author[Rossi et al.]{Elena M. Rossi$^{1,}$\thanks{E-mail: emr@strw.leidenuniv.nl}, T. Marchetti$^{1}$, M. Cacciato$^{1}$, M. Kuiack$^{2}$ and R. Sari$^{3,4}$ \\
$^{1}$Leiden Observatory, Leiden University, PO Box 9513, 2300 RA, Leiden, the Netherlands \\
$^{2}$ Anton Pannekoek Institute, University of Amsterdam, PO Box 94249, 1090 GE Amsterdam, The Netherlands\\
$^{3}$Racah Institute of Physics, Hebrew University, Jerusalem, Israel, 91904 \\
$^{4}$Theoretical astrophysics 350-17, California Institute of Technology, Pasadena, CA, 91125 \\
}

\begin{document}


\pagerange{\pageref{firstpage}--\pageref{lastpage}} \pubyear{2016}

\maketitle

\label{firstpage}


\begin{abstract}
  The mass assembly history of the Milky Way can inform both theory of
  galaxy formation and the underlying cosmological model. Thus,
  observational constraints on the properties of both its baryonic and
  dark matter contents are sought. Here we show that hypervelocity
  stars (HVSs) can in principle provide such constraints. We model the
  observed velocity distribution of HVSs, produced by tidal break-up
  of stellar binaries caused by Sgr A*. Considering a Galactic
    Centre (GC) binary population consistent with that inferred in
    more observationally accessible regions, a fit to current HVS data
    with significance level $> 5\%$ can {\it only} be obtained if the
    escape velocity from the GC to 50 kpc is $V_{\rm G} \lsim 850$
    \kms, regardless of the enclosed mass distribution. When a NFW
    matter density profile for the dark matter halo is assumed, haloes
    with $V_{\rm G} \lsim 850$ \kms are in agreement with predictions
    in the $\Lambda$CDM model and that a subset of models around
    $M_{\rm 200} \sim 0.5-1.5 \times 10^{12} M_{\odot}$ and
    $r_{\rm s} \lsim 35$ kpc can also reproduce Galactic circular
    velocity data. HVS data {\it alone} cannot currently exclude
    potentials with $V_{\rm G} > 850$ \kms. Finally, specific
    constraints on the halo mass from HVS data are highly dependent on
    the assumed baryonic mass potentials. This first attempt to
  simultaneously constrain GC and dark halo properties is primarily
  hampered by the paucity and quality of data. It nevertheless
  demonstrates the potential of our method, that may be fully realised
  with the ESA Gaia mission.
\end{abstract}



\section{Introduction}
The visible part of galaxies is concentrated in the centre of more
extended and more massive dark matter structures, that are termed
haloes. In our Galaxy, the baryonic matter makes up a few percent of
the total mass, and the halo is $\sim 10$ times more extended than the
Galactic disc. In the current paradigm, galaxies assemble in a
hierarchical fashion from smaller structures and the result is due to
a combination of merger history, the underlying cosmological model and
baryonic physics (e.g. cooling and star formation). Thanks to our
vantage point, these fundamental ingredients in galaxy assembly, can
be uniquely constrained by observations of the matter content of the Milky Way and its
distribution, when analysed in synergy with dedicated
cosmological simulations.

Currently, our knowledge of the Galactic dark matter halo is
fragmented. Beyond $\sim 10$ kpc dynamical tracers such as halo field
stars and stellar streams become rarer and rarer and astrometric
errors significant.  In particular, there is a large
uncertainty in the matter density profile, global shape, orientation
coarseness \citep[e.g.][]{bullock+10,L&M10,V&H13,loebman+14,laevens+15,W&E15} and
current estimates of the halo mass differ by approximately a factor
of 3 \citep[see fig.1 in][and references therein]{wang+15}.
This difference is significant as a mass measurement in the upper part
of that range together with observations of Milky Way satellites can
challenge \citep{klypin+99,moore+99,boylan+11} the current concordance cosmological paradigm: 
the so-called $\Lambda$ cold dark matter
model ($\Lambda$CDM).
In particular, the ``too big to
fail problem" \citep{boylan+11} states that, in $\Lambda$CDM high mass
($\gsim 2 \times 10^{12} M_{\odot}$) haloes, the most massive subhaloes
are too dense to correspond to any of the known satellites of the
Milky Way. Therefore, the solution may simply be a lighter Galactic halo of
$< 10^{12} M_{\odot}$ \citep[e.g.][]{vera+13,gibbons+14}. This is an
example of how a robust measurement of the Galactic mass can be
instrumental to test cosmological models.

On the other extreme of Galactic scales, the Galactic Centre (GC) has been
the focus of intense research since the beginning of the 1990s, and it
is regarded as a unique laboratory to understand the interplay between
(quiescent) supermassive black holes (SMBHs) and their
environment  \citep[see][for a review]{genzel+10}. Indeed, the GC harbours the best
observationally constrained SMBH, called Sgr A*, of mass
$\approx 4.0 \times 10^{6} M_{\odot}$ \citep{ghez+08,gillessen+09,meyer+12}. In
particular, GC observations raise issues on the
stellar mass assembly, which is intimately related to the SMBH growth
history. For example, in the central $r \sim 0.5$ pc the light is
dominated by young ($\sim 6$ Myr old) stars
\citep[e.g.][]{paumard+06,lu+13} with a suggested top-heavy initial mass function
\citep[IMF][]{bartko+10,lu+13} and a large spread in metallicity at $r<1$
pc \citep{do+15}. The existence of young stars well within the gravitational sphere
of influence of Sgr A* challenges our knowledge of how stars form, as
molecular clouds should not survive tidal forces there. These stars
are part of a larger scale structure called nuclear star cluster with
half-light radius around $\sim 5$ pc
\citep[e.g.][]{schodel+14,fritz+16}: in contrast with the inner
region, its IMF may be consistent with a Chabrier/Kroupa
IMF and between $2.5$ pc $<r<4$ pc the majority of
stars appear to be older than 5 Gyr
\citep[e.g.][]{pfuhl+11,fritz+16}. The origin of this nuclear star
cluster and its above mentioned features is highly debated, and the
leading models consider coalescence of stellar clusters that reach the
GC and are tidally disrupted or in situ formation from gas streams
\citep[see][for a review on nuclear star cluster]{boeker10}. The Hubble Space Telescope
imaging surveys have shown that most galaxies contain nuclear clusters
in their photometric and dynamical centres
\citep[e.g.][]{carollo+97,georvievboeker14,carson+15}, but the more
observationally accessible and best studied one is the Milky Way's,
which once more give us a chance of understanding the formation of
galactic nuclei in general. However, to investigate the GC via direct observations, one must cope with observational
challenges such as the strong and spatially highly variable
interstellar extinction and stellar crowding.
A  concise review of  the current knowledge of 
the nuclear star cluster at the GC and the observational obstacles and limitations is given in
\citet{schodel+14rev}.

Remarkably, a single class of objects can potentially address the mass
content issue from the GC to the halo: hypervelocity
stars (HVSs). These are detected in the outer halo \citep[but
note][]{zheng+14} with radial velocities exceeding the Galactic escape
speed \citep[Brown et al. 2005; see][for a review]{brown15}. So far around 20 HVSs have
been discovered with velocities in the range $\sim 300-700$ \kms,
and trajectories consistent with coming from the GC.
Because of the discovery strategy, they are all B-type stars mostly
in the masses range between $2.5-4 M_{\odot}$
\citep[e.g.][]{brown+14}. Studying HVSs is thus a
complementary way to investigate the GC stellar
population, by surveying more accessible parts of the sky. After
ejection, HVS dynamics is set by the Galactic gravitational
field. Therefore, regardless of their origin, HVS spatial and velocity
distributions can in principle probe the Galactic total matter distribution
\citep{gnedin+05,gnedin+10,yu&madau07,sesana+07,perets+09,fragione&loeb16}.

Retaining hundreds of \kms in the halo while originating from a deep
potential well requires initial velocities in excess of several hundreds of
\kms \cite{kenyon+08}, which are very rarely attained by stellar interaction mechanisms
put forward to explain runaway stars
\citep[e.g.][]{blaauw61,aarseth74,eldridge+11,perets12,tauris15,rimoldi+16}. 
Velocity and spatial distributions of runaway and HVSs are indeed expected to be different \citep{kenyon+14}. 
For example, high velocity runaway stars would almost exclusively come from the Galactic disc \citep{bromley+09}.
Instead, HVS energetics and trajectories strongly support the view that HVSs
were ejected in gravitational interactions that tap the gravitational
potential of Sgr A*, and, as a consequence of a huge ``kick'', escaped
into the halo. In particular, most observations are consistent with
the so called ``Hills' mechanism'', where a stellar binary is tidally
disrupted by Sgr A*. As a consequence, a star can be ejected with a
velocity up to thousands \kms \citep{hills88}. Another appealing feature is that
the observed B-type stellar population in the inner parsec --- whose in situ origin is quite unlikely --- 
is consistent with being HVSs' companions, left bound to Sgr A* by the Hills' mechanism
\citep{zhang+13,madigan+14}.

 In a series of
three papers, we have built up a solid and efficient semi-analytical
method that fully reproduces 3-body simulation results for mass ratios between 
a binary star and a SMBH
($m_{\rm t}/M \sim 10^{-6}$) expected in the GC. In particular we reproduce
star trajectories, energies after the encounter and ejection velocity
distributions \citep[see][and section \ref{sec: vej} in this
paper]{skr10,kobayashi+12,rossi+14}. Here, we will capitalise on that
work and apply our method to the modelling of current HVS data, with
the primary aim of constraining the Galactic dark matter halo and
simultaneously derive consequences for the binary population in the
GC. Since star binarity is observed to be very frequent
in the Galaxy (around $50\%$) and the GC seems no exception \citep[$\sim 30\%$ 
for massive binaries][]{pfuhl+14}, clues from HVS
modelling are a complementary way to understand the
stellar population within the inner few parsecs from Sgr A*.

This paper is organised as follows.  In Section \ref{sec: vej}, we
describe our method to build HVS ejection velocity distributions,
based on our previous work on the Hills' mechanism.  In Section
\ref{sec:1approach} , we present our first approach to predict
velocity distributions in the outer Galactic halo and we show our
results when comparing them to data in Section
\ref{sec:1approach_results}.  In Section \ref{sec:NFW}, we will
specialise to a ``Navarro, Frenk and White'' (NFW) dark matter profile and present results in Section
\ref{sec:second_results}. In Section \ref{discussion}, we discuss our findings, 
their limitations and implications and then conclude. Finally, 
in Appendix \ref{mcmc_appendix}, we describe our  analysis of the Galactic circular velocity data, 
that we combine with HVS constraints.


\section{Ejection velocity distributions}
\label{sec: vej}
We here present our calculation of the
{\em ejection} velocity distribution of hypervelocity stars (i.e. the
velocity distribution at infinity with respect to the SMBH) via the Hills' mechanism.
We denote with $M$ Sgr A*'s mass, fixed to $M=4.0 \times 10^{6} M_{\odot}$.

Let us consider a stellar binary system with separation $a$, primary mass
$m_{\rm p}$, secondary mass $m_{\rm s}$, mass ratio
$q = m_{\rm s}/m_{\rm p} \leq 1$, total mass $m_{\rm s}+m_{\rm p} = m_{\rm t}$ and period $P$. 
If this binary is scattered into
the tidal sphere of Sgr A*, 
the expectation is that its centre of mass is on a nearly parabolic orbit, as 
its most likely place of origin is the neighbourhood of Sgr A*'s radius of influence. Indeed, this latter is 
$\sim 5$ orders of magnitude larger than the tidal radius, and therefore the binary's orbit must
 be almost radial to hit the tiny Sgr A*'s tidal sphere. 
On this orbit, the binary star has\footnote{In \cite{skr10}, we show that a binary star on a parabolic orbit has
  80\% chance of disruption, when considering prograde and
  retrograde orbits. Our (unpublished) calculations averaged
  over all orbital inclinations indicate a high percentage around
  $\sim 90 \%$.} $\sim 90\%$  probability to undertake an exchange reaction,
  where a star remains in a binary with the black hole, while the companion is ejected.
  In addition, we proved that 
  the ejection probability is independent of the stellar mass, when the centre of mass of the binary is on a
  parabolic orbit.
  This is different from the case of elliptical or hyperbolic orbits where the primary star, 
carrying most of the orbital energy, has
  a greater chance to be respectively captured or ejected.
  \citep{kobayashi+12}.
  
The ejected star has a
velocity at infinity, {\em in solely presence of the black hole potential},
equal to 
\be v_{\rm ej} = \sqrt{\frac{2\,G m_{\rm c}}{a}}
\left(\frac{M}{m_{\rm t}}\right)^{1/6},
\label{eq:vedj}
\ee 
\citep{skr10} where $m_{\rm c}$ is the mass of the binary companion star to the HVS and $G$ is the 
gravitational constant.
Rigorously, there is a numerical factor in front of the square root in (eq.~\ref{eq:vedj})
that depends on the binary-black hole encounter geometry. However,
this factor is $\sim 1$, when averaged over the binary's
phase\footnote{The binary's phase is the angle between the stars' separation and their centre
 of mass radial distance from Sgr A*, measured, for instance, at the tidal radius or at pericentre.}.
 Moreover, the velocity distributions obtained with
the full numerical integration of a binary's trajectory and those
obtained with (eq.~\ref{eq:vedj}) are almost indistinguishable
\citep{rossi+14}. Given these results and the simplicity of eq. \ref{eq:vedj}, it is
possible to predict {\em ejection}
velocity distributions, efficiently exploring a large range of the parameter
space in Galactic potentials, binary separations and stellar masses. This latter is the main 
advantage over methods using 3-body (or
N-body) simulations.

Since we are only considering binaries with
primaries' mass $\gsim 3 M_{\odot}$, we may consider observations of
B-type and O-type binary stars for guidance. Because of
the large distance and the extreme optical extinction, 
observations and studies of binaries in the inner GC 
are limited to a handful of very massive early-type binary stars \citep[e.g.][]{ott+99,pfuhl+14}
and X-ray binaries \citep[e.g.][]{muno+05}. 

For more reliable statistical inferences, 
we should turn to observations of more accessible regions in the Galaxy and in the Large Magellanic Cloud (LMC).
They suggest that a
power-law description of these distributions is reasonable. 
In the Solar neighbourhood, spectroscopic binaries with primary masses
between $1-5 M_{\odot}$ have a separation distribution, $f_{\rm a}$, that for short
periods can be both approximated by a $f_{\rm a} \propto a^{-1}$
({\"O}pik's law, i.e.
$f(\log_{10}{P}) \propto \left(\log_{10}P\right)^{\eta}$, with $\eta =0$) and a log
normal distribution in period with $\left< P \right> \simeq 10$ day
and a $\sigma_{\rm logP} \simeq 2.3$
\citep{kouwenhoven+07,D&K13}. However, in the small separation regime,
relevant for the
production of HVSs, the log normal distribution may also be
described by a power-law\footnote{This fit value does not
  significantly depends on the total mass assumed for binaries. We do
  not calculate errors on this fitted index, because our aim is to
  draw in the $\gamma-\alpha$ parameter space an indicative range of 
  power-law exponents for the separation distribution of B-type
  binaries in the Solar Neighbourhood (see Figure \ref{fig:vGs}).}:
$f_{\rm a} \propto a^{0.8}$. For primary masses $> 16 M_{\odot}$, 
\citet{sana+12} find a relatively higher frequency of short-period binaries in Galactic young clusters,
$\eta \approx -0.55$, but a combination of a pick at the smallest periods and a power-law
may be necessary to encompass all available observations \citep[see e.g.][]{D&K13}. 
For this range of massive stars ($\sim 20 M_{\odot}$), a similar power-law distribution  $\eta \approx -0.45$
is also consistent with a statistical description of O-type binaries in the
VLT-FLAMES Tarantula Survey of the star forming region 30 Doradus of
the LMC \citep{sana+13}. In the same region, a similar analysis for observed early ($\sim 10 M_{\odot}$) B-type binaries 
recovers instead an {\"O}pik's law \citep{dunstall+15}. 

Mass ratio distributions, $f_{\rm q}$, for Galactic binaries are generally observed
to be rather flat, regardless of the
primary's mass range \citep[e.g.][see their table
1]{sana+12,kobulnicky+14,D&K13}.  Differently, in the 30 Doradus star
forming region, the mass ratio distributions appear to be steeper, 
($f_{\rm q} \propto q^{\sim (-1)}$ in O-type banaries and $f_{\rm q} \propto q^{\sim (-3)}$ in early
B-type ones), suggesting a preference for pairing with lower-mass 
companions: still
a power-law may be fitted to data \citep{sana+13,dunstall+15}.

We therefore assume a binary separation distribution 
\be f_{\rm a}
\propto a^{\alpha},
\label{eq:fa}
\ee
where the minimum separation is taken to be the Roche-Lobe radius
 $a_{\rm min}= 2.5 \times \max[R_*, R_{\rm c}]$, where $R_*$ and $R_{\rm c}$ 
are the HVS's and the companion's radii, respectively. As a binary mass ratio distribution, we assume
\be
f_{\rm q} \propto q^{\gamma}, 
\label{eq:fq}
\ee
for $m_{\rm min} \le m_{\rm s} \le m_{\rm p}$. If not otherwise stated, $m_{\rm min} =0.1 
M_{\sun}$.

The mass of the primary star ($m_{\rm p} \gtrsim 3 M_{\odot}$) is
taken from an initial mass function, that needs to mirror the star
formation in the GC in the last $\sim 10^9$ yr. As
mentioned in our introduction, the stellar mass function is rather
uncertain and may be spatially dependent.  Observations of stars with
$M > 10 M_{\odot}$ within about $0.5$ pc from Sgr A* indicate a rather
top-heavy mass function with $f_{\rm m} \propto m_{\rm p}^{-1.7}$
\citep{lu+13}.  At larger radii observations of red giants (and the
lack of wealth of massive stars observed closer in) may instead point
towards a more canonical bottom-heavy mass function
\citep[e.g.][]{pfuhl+11,fritz+16}. Given these uncertainties, we
explore the consequences of assuming either a Kroupa mass function \citep{kroupa02},
$f_{\rm m} \propto m_{\rm p}^{-2.3}$  or
top-heavy distribution, $f_{\rm m} \propto m_{\rm p}^{-1.7}$, in the
mass range $2.5 M_{\odot} \le m_{\rm p} \le 100 M_{\odot}$.

Finally, we do not introduce here any specific model for the injection
of binaries in the black hole tidal sphere and consequently, we do not explicitly consider any
``filter" or modification to the binary ``natal'' distributions. Likewise, we do not explicitly account for 
higher order multiplicity (e.g. binary with a third companion, i.e. triples) that may result in disruption of binaries with different distributions than those cited above.
On the other hand, a way to interpret
our results is to consider that the separation and mass ratio
distributions already contain those modifications. We
will explore these possibilities in Section~\ref{discussion}.


\section{Predicting velocity distributions in the halo: first approach.} 
 \label{sec:1approach} 
 In this Section, we first describe how we
 compute {\em the halo} velocity distribution with a method that
 allows us to use a single parameter to describe the Galactic
 deceleration, without specifying its matter profile (Sec. \ref{sec:fv_vg}) . Given the large
 Galactocentric distances at which the current sample of HVSs is
 observed, our method is shown to be able to reproduce the correct
 velocity distribution for the velocity range of interest, {\em
   without} the need to calculate the HVS deceleration along the
 star's entire path from the GC. These features allow us
 to efficiently explore a large range of the binary
 population and the dark matter halo parameter space. Then, in
 Sec. \ref{sec:1approach_data_comp}, we describe how we perform our
 comparison with current selected data and finally we present our results in Sec. \ref{sec:1approach_results}.


\subsection{Velocity distribution in the halo: global description of the potential}
\label{sec:fv_vg}
Our first approach follows \cite{rossi+14} and consists in {\em not} assuming
any specific model for the Galactic potential, but rather to globally describe
it by the minimum velocity, $\vgs$, that an 
object must have at the GC in order to reach 50 kpc with a velocity equal or 
greater than zero. 
In other words, the parameter \vg is a measure of the net deceleration
suffered by a star ejected at the GC into the outer halo,
regardless of the mass distribution interior to it. The statement is that 
Galactic potentials with the same \vg produce the same velocity distribution beyond 50 kpc,
which is where most HVSs are currently observed\footnote{There is one discovered at $\sim 12$ kpc 
\citep{zheng+14}, but we will not include in our analysis because it has a different mass and location 
than our working sample, and therefore it would need a separate analysis.}.

The physical argument that supports this statement is the following. For any reasonable 
distribution of
mass that accounts for the presence of the observed bulge, most of the
deceleration occurs well before stars reach the inner halo \citep[e.g.][]{kenyon+08} and therefore, any
potential with the same escape velocity \vg will have the same
net effect on an initial ejection velocity:
 \be 
 v = \sqrt{v_{\rm ej}^2 - V_{\rm G}^2}. 
 \label{eq:v}
 \ee 
Although practically we are interested in the HVS distribution beyond $50$ 
kpc, the method outlined here is valid for any threshold distance as long as the deceleration beyond that is negligible and, 
as justified below, all stars in the velocity range of interest reach it within their life-time. 
Therefore in the following, when a specific choice is not needed, we will generically call this 
threshold distance ``$r_{\rm in}$". This, we recall, is also the radius associated to \vg.

Let us now proceed to calculate the HVS velocity distribution within a given radial range 
$\Delta r=[r_{\rm out}-r_{\rm in}]$ in spherical symmetry, assuming a time-independent ejection rate 
$\cal{R}$ (typically $\sim 10-100$ Myr$^{-1}$).
Given the above premises, HVSs with a velocity around $v$ cross $r_{\rm in}$ at a rate 
$d\dot{N}/dv$, that can be obtained from the ejection-velocity probability density function (PDF) 
$P(v_{\rm ej})$ equating bins of corresponding velocity,
$$ \frac{d\dot{N}}{dv} dv = {\cal R} P(v_{\rm ej}) {\rm d}v_{\rm ej}, $$ with the aid of eq.\ref{eq:v}, that gives $v=v(v_{\rm ej})$.
Consequently, the halo-velocity PDF ($dn/dv$) within a given radial range $\Delta r$ can be simply computed as
\be
dn(v,\Delta r) \propto  \frac{d\dot{N}}{dv} \times \min[ \Delta r/v, \left<t_{\rm life}\right>] \,dv,
\label{eq:dnv_VG}
\ee 
where $\min[ \Delta r/v, \left<t_{\rm life}\right>]$ is the average residence time in that range of Galactocentric distances 
of HVSs in a bin  $dv$ of velocity around $v$. This is the minimum between the crossing time $\Delta r/v$ and the average life-time $\left<t_{\rm life}\right>$
beyond $r_{\rm in}$ of a star in that velocity bin. This latter
term accounts for the possibility that  stars may evolve out of the
main sequence and meet their final stellar stages before they reach
the maximum radial distance considered (i.e. $r_{\rm out}$) . 

More precisely for a given star $t_{\rm life}$ should be equal
  to the time left from its main sequence lifetime $t_{\rm MS}$, after
  it has dwelled for a time $t_{\rm ej}$ in the GC, and subsequently
  travelled to $r_{\rm in}$ in a flight-time $\tau(r_{\rm in})$:
  $t_{\rm life} =t_{\rm MS} -(t_{\rm ej}+\tau(r_{\rm in}))$.
  Observations suggest that a HVS can be ejected at anytime during its
  lifetime with equal probability and therefore on
  average $t_{\rm ej} \approx t_{\rm MS}/2$ \citep{brown+14}. In addition, if
  $\tau(r_{\rm in}) \ll t_{\rm MS}$, we can write
  $\left<t_{\rm life} \right> = \left<t_{\rm MS}\right>/2$, where
  $\left<t_{\rm MS}\right> = \int (dn/dm) ~t_{\rm MS}(m) dm$ is the
  average main sequence life-time weighted for the star mass
  distribution $dn/dm$ in a given velocity bin. 
  
  In the HVS mass and
  metallicity range considered here $t_{\rm MS}(m) \approx 200-700$ Myr (and 
  $\left<t_{\rm MS}\right> \approx 300 - 600$ Myr). 
  Consequently our calculations typically show
  $\tau(r_{\rm in}) < t_{\rm MS}$ for velocities $> 150$ \kms,
  when adopting $r_{\rm in} =50$ kpc. This
  means that $\tau(r_{\rm in}) \ll t_{\rm MS}$ in the whole velocity
  range of interest in this work ($v \geq 275$ \kms, see Section
  \ref{sec:1approach_data_comp}).

In this framework, we construct a Monte Carlo code where $10^{7}$
binaries are drawn from the distributions described in Section
\ref{sec: vej} to build an ejection velocity PDF. This is used to
construct the expected PDF in the outer halo (eq.\ref{eq:dnv_VG})
between $r_{\rm in}=50$ kpc and $r_{\rm out}=120$ kpc (the observed
radial range), using the formalism detailed above. For each bin of
velocity, we calculate the $\left<t_{\rm MS}\right>$, using the
analytical formula by \citet[][see their equation 5]{hurley+00}. The
lifetime for a star in the $2.3-4 M_{\odot}$ range is of a few to
several hundred million years, but the exact value depends on
metallicity (higher metallicities correspond to longer
lifetimes).  Until recently, solar metallicity was thought to be the
typical value for the GC stellar population. However,
more recent works suggest that there is a wider spread in metallicity,
with a hint for a super-solar mean value \citep{do+15}. 
\begin{figure*}
\centering
\includegraphics[width=\textwidth]{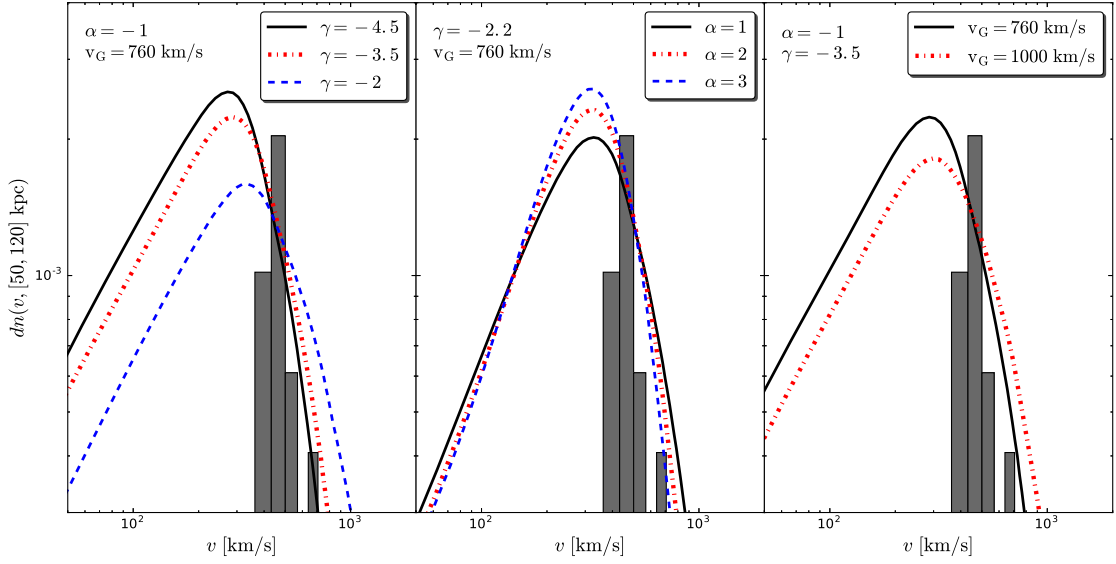}
\caption{Probability density functions for HVS velocities in the outer
  halo of our Galaxy, between 50 kpc and 120 kpc.  They are calculated
  following the deceleration procedure explained in
  Section~\ref{sec:1approach} and depend on 3 main parameters:
  $\gamma$, $\alpha$ (for the binary mass ratio and semi-major axis
  distributions) and \vg.  In each panel, two parameters are kept
  fixed while we show how the distribution changes by changing the
  value of the third parameter. See text for a detailed description.
  For a visual comparison, we over-plot data from \citet{brown+14}
  (``unbound sample'' only), with an arbitrary binning.}
\label{fig:PDF}
\end{figure*}
In the following, our fiducial model will assume:
\begin{itemize}
\item HVSs  masses between 2.5 and 4 solar masses;
\item A Kroupa ($f_{\rm m} \propto m_{\rm p}^{-2.3}$) IMF for primary
  stars between 2.5 and 100 solar masses;
\item For a given primary mass $m_{\rm p}$, a mass ratio distribution
  $f_{\rm q} \propto q^{\gamma}$ in the range
  $[m_{\rm min}/m_{\rm p}, 1]$, with $m_{\rm min}=0.1M_{\odot}$ and
  $-10 \le \gamma \le 10$;
\item A separation distribution $f_{\rm a} \propto a^{\alpha}$ between
  $a_{\rm min}=2.5 \times \max[R_*, R_{\rm c}]$ and
  $a_{\rm max}=10^3 R_{\odot}$, with $-10 \le \alpha \le 10$;
\item A HVS mean metallicity value of $Z=0.05$ (i.e. super-solar). 
\end{itemize} 
We will explore different assumptions in Section~\ref{discussion}. In
particular, we will investigate a top-heavy primary IMF, explore the
consequence of a solar metallicity and finally assume a higher value
of $m_{\rm min}$, over which we have no observational constraints in
the GC. We will find that only the latter, if physically possible, may
significantly impact our results and will discuss the consequences.

Examples of velocity distributions in the halo for our fiducial model
are shown in Figure \ref{fig:PDF}.  Our selected data (see the
Figure's caption and next Section) are over-plotted with an arbitrary
binning (histogram). It is here worth reminding some of the features
derived in \citet{rossi+14}. There, we analytically and numerically
showed that the HVS halo velocity distribution encodes different
physical information in different parts of the distribution.  In
particular, the peak of the distribution depends on both \vg and the
binary distributions, and moves towards lower velocity for lower \vg
(right panel) and higher values of $|\gamma|$ and $\alpha$ (left and
central panels).  On the other hand, the high-velocity branch only
depends on the binary properties, as the Galactic deceleration is
negligible at those velocities. From eq.\ref{eq:dnv_VG}, one can
derive that for $v \gg v_{\rm G}$ the high-velocity branch is
independent of the binary semi-major axis distribution (i.e. $\alpha$)
for $\gamma > -(\alpha+2)$ and
$$ dn \propto v^{2\gamma} dv.$$ Therefore larger value of $|\gamma|$ result in a steeper distribution at high velocities. 
This is shown in the left panel of Figure \ref{fig:PDF}.  Instead in
the $v \gg v_{\rm G}$ and $\gamma < -(\alpha+2)$ regime,
$$ dn \propto v^{-2(\alpha+2)} dv,$$ independently of the assumed mass
ratio distribution and a steeper power-law is obtained for larger
$\alpha$ values (central panel).  A discussion on the low-velocity
tail, that it is {\em solely} shaped by the deceleration, is postponed
to Section \ref{sec:low_velocity_tail}.

\subsection{Comparison with data}
\label{sec:1approach_data_comp}
Beside the current HVS sample of so-called ``unbound'' HVSs (velocity
  in the standard rest frame $\gsim 275$ km s$^{-1}$), there is an
equal number of lower velocity ``bound'' HVSs\footnote{Here, we simply
  follow the nomenclature given in \citet{brown+14} of the two
  samples, even if, in fact, a knowledge of the potential is required
  to determine whether a star is bound and this is what we are
  after.}.  Currently, it is unclear if they all share the same origin
as the unbound sample, as a large contamination from halo stars cannot
be excluded.  We will therefore restrict our statistical comparison
with data to the unbound sample \citep[see upper part of table
1in][]{brown+14}.  As mentioned earlier, we only select HVS with
masses between $2.5-4 M_{\odot}$, with Galactocentric distances
between 50 kpc and 120 kpc, for a total of 21 stars.  These selections
in velocity, mass and distance will be also applied to our predicted
distributions.

Specifically, we calculate the total PDF as described by
eq.~\ref{eq:dnv_VG} and we perform a one dimensional
Kolmogorov-Smirnov (K-S) test applied to a left-truncated data
sample\footnote{See for example: Chernobai, A., Rachev, S. T., and
  Fabozzi, F. J. (2005).  Composite goodness-of-fit tests for
  left-truncated loss samples. Technical Report, University of
  California, Santa Barbara}. If we call $n(<v, \Delta r)$ the
cumulative probability function (CPF) for HVS velocities in the
distance range $\Delta r$, then the {\em actual CPF} that should be
compared with data is, \be n^{*}(<v, \Delta r) = \frac{n(<v, \Delta
  r)- n(< 275~{\rm km ~s^{-1}}, \Delta r)}{1- n(< 275~{\rm km
    ~s^{-1}}, \Delta r)}.  \ee Therefore, the K-S test result is
computed as \be D \equiv \max[| n^{*}(<v, \Delta r) - n_{\rm d}(<v)|],
\ee where $n_{\rm d}(<v)$ is the CPF of the actual data The
significance level $\bar{\alpha} = 1 - P(D \le \bar{d})$ is the
probability of rejecting a fitted distribution $n(<v, \Delta r)$, when
in fact it is a good fit.  The most commonly used threshold levels for
an acceptable fit are $\bar{\alpha}=0.01$ and $\bar{\alpha}=0.05$.
For 21 data points $\bar{d}=0.344$ and $\bar{d}=0.287$ are the
critical values below which the null hypothesis that the data are
drawn from the model cannot be rejected at a significance level of 1\%
and 5\% respectively.

Note that no HVS is observed with a velocity in excess of $v > 700 $
km s$^{-1}$. Since the HVS discovery method is spectroscopic as
opposed to astrometric, there is no obvious observational bias that
would have prevented us from observing HVS with $v > 700 $ \kms within
120 kpc and so we do not perform any high-velocity cut to our
model\footnote{We remark in addition that our eq.~\ref{eq:dnv_VG}
  takes already into account that faster stars have a shorter
  residence time by suppressing their number proportionally to
  $v^{-1}$}.  Indeed, the absence of high-velocity HVSs in the current
(small) sample suggests that they are rare, and this fact puts strong
constraints on the model parameters.  From the discussion in the
previous section, a suppression of the high-velocity branch can be
achieved by either choose a lower \vg or choose steeper binary distributions
(a larger $|\gamma|$ or $\alpha$), as we will explicitly show in the
next section.

\begin{figure*}
\includegraphics[width=\textwidth]{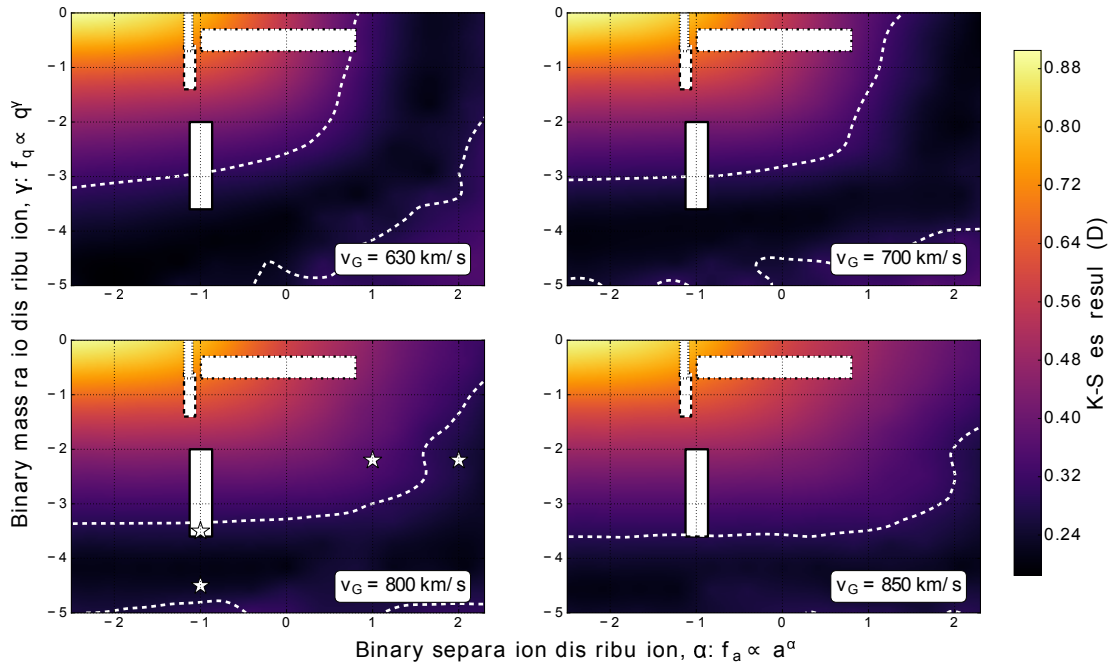}
\caption{Contour plots for K-S test results in the parameter space
  $\alpha-\gamma$ for 4 different values of $V_{\rm G}$ (see panels'
  label). The white dashed line indicates the $5\%$ significance level contours.
  The white regions correspond to
  observed properties of B-type or O-type binaries: the region
  enclosed by a dash-dotted line is for late B-type stars ($2-5 M_{\odot}$)
  in the Solar Neighbourhood \citep{kouwenhoven+07,D&K13}; results for Galactic O-type binaries are shown within the
  region marked by a dotted line \citep{sana+12}; the region
  enclosed by a solid (dashed) line is for early $\sim 10 M_{\odot}$ B-type (O-type) binaries
  observed in 30 Doradus \citep{sana+13,dunstall+15}. The
  four stars mark the points ($\alpha, \gamma$) in the
  parameter space for which the PDF is shown in Figure \ref{fig:PDF}
  (see also Fig.\ref{fig:Mh_rs_KS}).}
\label{fig:vGs}
\end{figure*}

\subsection{Results} 
\label{sec:1approach_results} 
In each panel of Figure~\ref{fig:vGs}, we explore the parameter space
$\alpha-\gamma$ for a fixed global deceleration that brakes stars
while travelling to 50 kpc, i.e. for a given $\vgs$. The contour plots
show our K-S test results and models below and at the right of the
white dashed line have a significance level higher than 5\%:
i.e. around and below that line current data are consistent with
coming from models with those sets of parameters.

Let us first focus on the upper right panel  ($\vgs \approx 700 $
\kms), as it shows clearly a common feature of all our contour
plots in this parameter space. There is a stripe of minima
that, from left to right, first runs parallel to the $\alpha$-axis and
then to the $\gamma$-axis\footnote{We note that, even if not
  completely apparent in all our panels, the K-S test values start to
  increase again moving towards high values of $|\gamma|$ and
  $\alpha$: i.e. the stripe of minima has a finite size.}.  This
stripe is the locus of points where the high-velocity tail of the
distributions has a similar slope: this happens for values of  
$\gamma$ and $\alpha$ related by
$\gamma \approx -(\alpha+2)$ (see discussion of Figure \ref{fig:PDF} in Section \ref{sec:fv_vg}).
For negative $\alpha$
values (distributions with more tight binaries than wide ones), the
high-velocity distribution branch is mainly shaped by the mass ratio distribution
and, for example in this panel, a value around $\gamma \approx -4$
gives the best fit. On the other hand, for positive $\alpha$
(i.e. more wider binaries than tight ones), the high-velocity tail is
shaped by the separation distribution and a value of around
$\alpha \approx 2$ gives the best K-S results. 

When increasing the escape velocity (from top left to bottom right)
the stripe of minima moves towards the right lower part of the plots
and gets further and further from the regions in the $\alpha-\gamma$
parameter space that correspond to observations of B-type binaries,
and actually, to our knowledge, of {\em any} type of binaries
currently observed with enough statistics in both star-forming and
quiescent regions. We focus on observations of B-type binaries
because, although our calculation consider $\sim 3 M_{\odot}$ HVSs
ejected from binaries with all possible mass combinations, we find
that the overall velocity distribution is highly dominated by binaries
where HVSs were the primary (more massive) stars, i.e. late B-type
binaries\footnote{Binaries where the HVS companions are the primary
  stars just contribute at a percentage level and only to the highest
  velocity part of the velocity distribution (see eq.\ref{eq:vedj}) in
  the whole parameter space explored in this work.}. 
     
In all panels, but the bottom right one, the white dashed line crosses
or grazes the $\alpha-\gamma$ parameter space indicated by a white
rectangle within a solid black line. We conclude that within an
  approximate range $ V_{\rm G} \lsim 850 $ \kms, the current
observed HVS velocity distribution can be explained assuming a binary
statistical description in the GC that is consistent with the one
inferred by \citet{dunstall+15} for $\sim 10 M_{\odot}$ B-type
binaries in the star forming region of the Tarantula Nebula. In
  addition, for $V_{\rm G} \lsim 630$ \kms the 5\% confidence line
  {\it also} crosses the parameter space observed for Galactic B-type
  binaries \citep{kouwenhoven+07}.  An argument in favour of a
similarity between known star forming regions and the inner GC is
that, in this latter, \citet{pfuhl+14} infer a binary fraction close
to that in known young clusters of comparable age.  However, we warn
the reader that the Tarantula Nebula's results are affected by
uncertainties beyond those represented by the nominal errors on
$\alpha$ and $\gamma$ reported by \citet{dunstall+15} and we will
discuss those in Section~\ref{discussion}.

Finally, we comment on our choice to define
  the \vg limit using a 5\% significance level threshold. If we relax
  this assumption and accept models with significance level $> 1\%$
  (another commonly used threshold) the \vg limit moves up to
  $\vgs \approx 930$ \kms. On the other hand, models with $>10\%$
  significance level have $\vgs \lsim 800$ \kms. Therefore, as a representative value,
  we cite here and thereafter the intermediate one of $850$ \kms, corresponding to the 5\% 
threshold.

\begin{figure}
\includegraphics[width=0.5\textwidth]{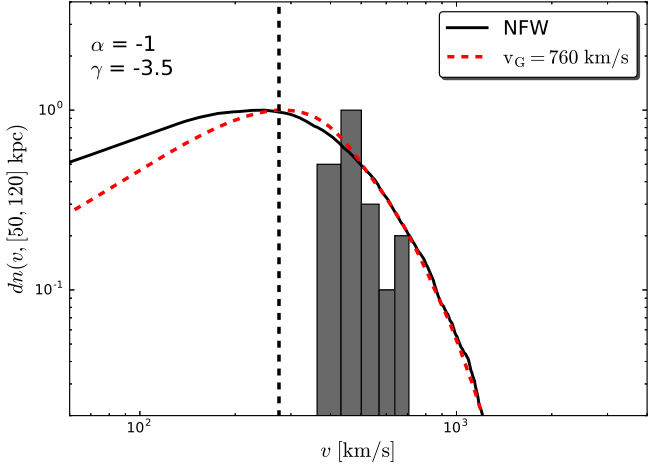}
\caption{ Galactic halo velocity distributions between 50 and 120 kpc for a fixed binary
  statistical description (see parameters in the upper left corner)
  but with different treatments of the star deceleration: the red
  dashed line is computed as described in Section \ref{sec:fv_vg} for
  $\vgs = 760$ \kms while the black solid line is our model where
  stars are continuously decelerated in a potential whose halo is
  described by a NFW profile with mass
  $M_{\rm h} = 0.5 \times 10^{12} M_{\odot}$ and scale radius
  $\rss = 31$ kpc (see Section~\ref{sec:NFW}). This potential requires
  an initial velocity to escape from the GC to 50 kpc of
  $V_{\rm G} \approx 760$ \kms (see eq. \ref{eq:vg_NFW}). Unlike
    Figure \ref{fig:PDF}, both model distributions and data are
    normalised at the peak for an easier visual comparison. The
    vertical dashed line marks the selection threshold ($v=275$ \kms)
    of the Brown et al. unbound sample. This comparison shows
    that for $v \gsim 250$ \kms the two distributions are similar, as
  confirmed by the results from the K-S test ($D = 0.25$ for the black
  solid line and $D = 0.26$ for the red dashed line).}
\label{fig:3}
\end{figure}

\section{Second approach: assuming a Galactic Potential model} \label{sec:NFW}
 
We now choose a specific model to describe the Galactic potential, in order to cast our results 
in terms of 
dark matter mass and its spatial distribution.

We represent the dark matter halo of our Galaxy with a Navarro Frank and White (NFW) profile,
\be
\phi(r)_{\rm NFW} =- G M_{\rm h} \left(\frac{\ln(1+ r/r_{\rm s})}{r}\right),
\label{eq:nfw}
\ee \citep{nfw96}. In this spherical representation there are only two
parameters: the halo mass $M_{\rm h}$ and the scale radius
$r_{\rm s}$, where the radial dependence changes.  Eq.\ref{eq:nfw}
assumes an infinite potential (no outer radius truncation) which is
justified in our case since we consider Galactocentric distances
smaller than the halo virial radius ($\sim 200$ kpc).

The baryonic mass components of the Galactic potential can be described by a Hernquist's 
spheroid for the
bulge \citep{hernquist90},
\be \phi(r)_{\rm b} = -\frac{G M_{\rm b}}{r+r_{\rm b}}, 
\label{eq_phi_b}
\ee 
(in spherical coordinates) plus a Miyamoto-Nagai disc \citep[][in cylindrical coordinates, where $r^2=R^2+z^2$]{M&N75}, 
\be 
\phi_{\rm d} (R,z) = - \frac{G M_{\rm d}} {\sqrt{R^2 +\left(a+\sqrt{z^2+b^2} \right)^2}},
\label{eq_phi_d}
\ee with the following parameters:
$M_{\rm b} = 3.4 \times 10^{10} M_{\odot}$, $r_{\rm b} =0.7$ kpc,
$M_{\rm d} = 1.0 \times 10^{11} M_{\odot}$, $a=6.5$ kpc and $b=0.26$
kpc. This Galactic model have been used in modelling both HVSs and
stellar streams \cite[e.g.][and with slightly different parameters by
Kenyon et al. 2008]{johnston+95,Price2014,hawkins+15}.
Observationally, our choice for the bulge's mass profile is supported
by the fact that its density profile is very similar to that obtained
by \cite{kafle+14}, fitting kinematic data of halo stars in SEGUE\footnote{The
\cite{kafle+14} model for the bulge is not spherical (see their table 1), therefore we
  compare to our model both their spherically averaged density profile
  and their density profile at $45^{\circ}$ latitude (see Section
  \ref{sec:NFW} for a justification of this latter).}. In addition \cite{kafle+14} use our same model for the disc
mass distribution and their best
fitting parameters are very similar to our parameters (see their table
1 and 2). However, different choices may also be consistent with
current data, and we will discuss the
impact of different baryonic potentials on our results in Section
\ref{sec:second_results}.

In a potential constituted by the sum of all Galactic components, 
\begin{equation}\label{eq:sumofpotentials}
\phi_{\rm T} (r, M_{\rm h}, r_{\rm s}) = \phi(r(R,z))_{\rm d} +\phi(r)_{\rm b} + \phi(r)_{\rm NFW} \, ,
\end{equation}
we integrate each star's trajectory from an inner radius
$r_{\rm start}=3$ pc, equal to Sgr A*'s sphere of influence but  any starting radius
$r_{\rm start}< 20$ pc gives very similar results.
 In fact, we find that the disc's
sky-averaged deceleration is overall negligible with respect to that
due to the bulge. To save computational time, we therefore set
$R=z=r/\sqrt{2}$ in equation \ref{eq_phi_d} (i.e. we only consider
trajectories with a Galactic latitude of 45$^{\circ}$), simplifying
our calculations to one-dimensional (the Galactocentric distance $r$)
solutions.

The star's initial velocity is drawn from the ejection velocity
distribution, constructed as detailed in Section \ref{sec:
  vej}. Assumptions on HVS properties are those of our fiducial
model. Informed by observations \citep{brown+14}, we assigned a
flight-time from a flat distribution between $[0, t_{\rm MS}]$.  Each
integration of $10^7$ star orbits gives a sky realisation of the
velocity PDF, but we actually find that the number of stars we are
tracking is sufficiently high that differences between PDFs associated
to different realisations are negligible.

An example of a halo velocity distribution is shown in Figure
\ref{fig:3} with a black solid line. This accurate calculation of the
star deceleration is well approximated by using eq.\ref{eq:v} for
  $v \gsim 250$ \kms, when the escape velocity at 50 kpc is
calculated as \be V_{\rm G}^2 = 2 (\phi_{\rm T} \left(50 ~{\rm kpc},
  M_{\rm h}, r_{\rm s}) - \phi_{\rm T} (r_{\rm start}, M_{\rm h},
  r_{\rm s})\right),
\label{eq:vg_NFW}
\ee (red dashed line in Figure~\ref{fig:3}). Despite the
  discrepancy in the behaviour of the low velocity tail, the two
approaches give very similar K-S test results when compared to current
observations ($D =0.26$ for the NFW model versus $D =0.25$ for the
``\vg'' model). With a random sampling, we tested that K-S results
differ at most at percentage level in the whole extent of the
parameter space of interest to us, validating our first approach, as
an efficient and reliable exploratory method.

\subsection{The low-velocity tail}
\label{sec:low_velocity_tail}
We here pause to discuss and explain the difference in the velocity
distribution around and below the peak calculated with our two
approaches (see Figure~\ref{fig:3}). Without loss of indispensable
information, the impatient reader may skip this section and proceed to
the next one, where we discuss our results.

The low velocity tail discrepancy is due to our two main assumptions
of our first method: i) neglecting the residual deceleration beyond
$50$ kpc; and ii) all stars reach $50$ kpc before they evolve out of
the main sequence. The residual deceleration gives an excess of low
velocity stars in the correct distribution (black solid line) that
cannot be reproduced by our approximated calculation (red dashed
line).  On the other hand, a fraction of stars that should have ended
up with velocities $ \lsim 150$ \kms beyond $50$ kpc have in fact
flight-times longer than their life-time and the low velocity excess
is slightly suppressed in that range.

Let us be more quantitative. In the framework of our first approach,
one can show that the PDF at low velocities increases linearly with
$v$ \citep{rossi+14}. The calculation is as follows. The rate of HVSs
crossing $r=r_{\rm in}$ with
$v =\sqrt{v_{\rm ej}^2-V_{\rm G}^2} \ll V_{\rm G}$ is given by
$$ \frac{d\dot{N}}{dv} \sim {\cal{R}} \left.P(v_{\rm ej})\right |_{v_{\rm ej}=V_{\rm G}} \frac{v}{V_{\rm G}}.$$ 
Moreover, for\footnote{We remind the reader that
  $\Delta r = r_{\rm out}-r_{\rm in}$.}
$$v < \Delta r / \left<t_{\rm MS}\right> \approx 230~{\rm km~s}^{-1}(\Delta r /70 \rm kpc)(300 /~\rm Myr/\left<t_{\rm MS}\right>),$$ the residence time within
$\Delta r$ is equal to (half of)
the stars' life-time, therefore from eq.\ref{eq:dnv_VG} we conclude
that
$$\frac{dn(v, \Delta r)}{dv} \propto \left. P(v_{\rm ej})\right |_{v_{\rm ej}=V_{\rm G}} v \times \left<t_{\rm MS}\right>,$$
recovering the linear dependence on $v$. In fact,
$\left<t_{\rm MS}\right>$ is not completely independent of $v$ as it
varies by a factor of $\approx 1.5$ as $v \rightarrow 0$. Therefore
$dn/dv$ is slightly sub-linear in $v$. The dependence of $\left<t_{\rm MS}\right>$ on $v$ comes about because 
$v_{\rm ej}$ is proportional to $m_{\rm c}$. This causes low-velocity
HVSs to be increasingly of lower masses
$(\rightarrow 2.5 M_{\odot})$, being ejected from binaries where their
companions were all lighter $m_{\rm c} \lsim 2.5 M_{\odot}$ than the companions of more massive HVSs.

When considering instead the full deceleration of stars in a
gravitational potential $a = -d\phi_{\rm T}(r)/dr$ as they travel
towards $r_{\rm out}$, their velocity depends both on $v_{\rm ej}$ and
$r$,
\be
v(v_{\rm ej},r)=\sqrt{v_{\rm ej}^2-\left(V_{\rm esc}(0)^2-V_{\rm esc}(r)^2\right)},
\label{eq:v_sec4.1}
\ee      
where $V_{\rm esc}(r)$ is the escape
velocity from a position r to infinity (i.e. $V_{\rm esc}(0)$ is the
escape velocity from the GC to infinity). Note that
$V_{\rm G} = \sqrt{V_{\rm esc}(0)^2-V_{\rm esc}(r_{\rm in})^2}$. In
the example shown in Figure~\ref{fig:3}, $V_{\rm esc}(0) \approx 826$
\kms, $V_{\rm esc}(r_{\rm in}=50 ~\rm kpc) \approx 323$ \kms,
$V_{\rm esc}(r_{\rm out}=120 ~\rm kpc) \approx 257$ \kms and
$V_{\rm G} \approx 760$ \kms. On the other hand, the distance $r$ is a
function of both $v_{\rm ej}$ and the flight-time
$\tau(r) = \int dv(r)/\left|a(r)\right|$, and this latter is a
preferable independent variable because uniformly
distributed. Therefore we express $v=v(v_{\rm ej},\tau)$ and 
\be
\frac{dn}{dv} \propto \int_{0}^{\left<t_{\rm MS}\right>}
\int_{v_{\rm ej,min}}^{v_{\rm ej,max}} \delta{(v - v(v_{\rm
    ej},\tau))} P(v_{\rm ej}) dv_{\rm ej} d\tau,
\label{eq:dn_dec}
\ee 
where the relevant ejection velocity range is that that gives
low-velocity stars between $r_{\rm in}$ and $r_{\rm out}$:
$v_{\rm ej,min} = \sqrt{v^2+\left(V_{\rm esc}(0)^2-V_{\rm esc}(r_{\rm
      in})^2\right)}$ and
$v_{\rm ej,max} = \sqrt{v^2+\left(V_{\rm esc}(0)^2-V_{\rm esc}(r_{\rm
      out})^2\right)}$. Note that, for Galactic mass distribution
where
$V_{\rm esc}(0) > V_{\rm esc}(r_{\rm in}),V_{\rm esc}(r_{\rm out})$,
the range $[v_{\rm ej,min}-v_{\rm ej,max}]$ is rather narrow and for
$v \ll V_{\rm G}$ these limits may be taken as independent of
$v$. This is the case in the example of Fig. \ref{fig:3}, where
$v_{\rm ej,min} \approx V_{\rm G} \approx 760 < v_{\rm ej} [\rm km
~s^{-1}] < v_{\rm ej,max} \approx 785$.

It follows that the low-velocity tail is populated by stars that where
ejected with velocities slightly higher than $V_{\rm G}$. If we
further assume that the flight-time $\tau$ to reach any radius within
$r_{\rm out}$ is always smaller than $\left<t_{\rm MS}\right>$
(formally this means putting the upper integration limit in $\tau$
equal to infinity), then all HVSs ejected with that velocity reach 50
kpc. It may be therefore intuitive that, applying the above
considerations, eq.\ref{eq:dn_dec} reduces to 
\be 
\frac{dn}{dv}(v,\Delta r) \propto \left.P(v_{\rm ej})\right |_{v_{\rm ej}=V_{\rm G}}
\int_{r_{\rm in}}^{r_{\rm out}} \frac{dr}{v_{\rm ej}(r)} \approx
\left.P(v_{\rm ej})\right |_{v_{\rm ej}=V_{\rm G}} \frac{\Delta r}{V_{\rm G}},
\label{eq:dn_dec_2}
\ee 
where we substitute $d\tau = dv/|a|$ in eq.\ref{eq:dn_dec} and we use eq.\ref{eq:v_sec4.1}.
We therefore recover the flat behaviour for $v \lsim 300$ \kms of
the black solid line in Figure \ref{fig:3}. We, however, also notice
that below $\sim 150$ \kms there is a deviation from a flat
distribution: this is because our assumption of
$\tau(r_{\rm in}) \ll \left<t_{\rm MS}\right>$ breaks down, as not all
stars reach $50$ kpc, causing a dearth of HVSs in that range.

As a concluding remark, we stress that, although we do not apply it
here, the result stated in eq.\ref{eq:dn_dec_2} can be used to further
improve our first method, a necessity when low-velocity data will be
available.

    \begin{figure*}
    \centering
    \includegraphics[width=\textwidth]{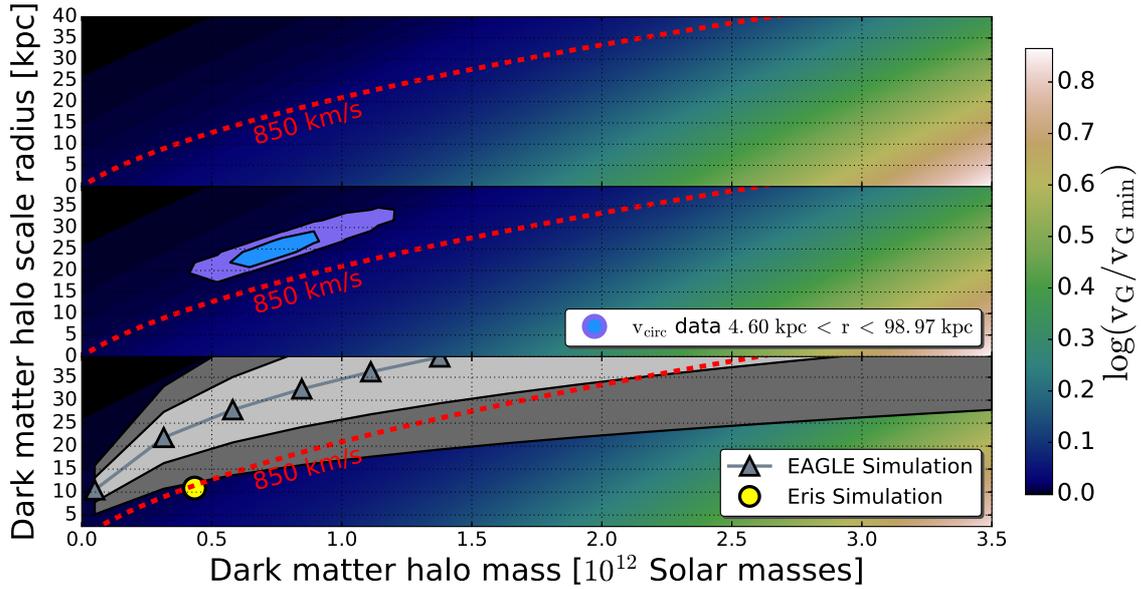}
    \caption{{\em Upper panel:} the ``escape" velocity from the GC to $50$ kpc, $\vgs$,
    over the minimum allowed by the presence of a baryonic disc and
    bulge ($V_{\rm G,min} = 725$ \kms)~is mapped onto the $\Mhs-\rss$
    parameter space for NFW dark halo profiles using
    eq. \ref{eq:vg_NFW}. The iso-contour line equal to $\vgs=850$ \kms
    is explicitly marked as red dashed line.
    {\em Middle panel:} same as the upper
    panel but over-plotted are the results of our MCMC 
    analysis of the Galactic circular velocity data from \citet{huang+16} (see Appendix
    \ref{mcmc_appendix}).
    {\em Lower panel:} the same as the upper
    panel but over-plotted are results from the 
    Eris 
    \citep{guedes+11} 
    and EAGLE 
    \citep{schaye+15}
     simulations. These are
    dark matter plus baryons simulations: the first one is a single
    realisation of a Milky Way-type galaxy, the latter are
    cosmological simulations that span a wider range of masses
    $(10^{10}-10^{14} M_{\odot})$. Following \citet{schaller+15},
    figure 11 middle panel, we plot the mass concentration relation
    found in EAGLE in our mass range, with a scatter in the
    concentration parameter of $25\%$ at one sigma level.}
    \label{fig:vg_mh_rs}
    \end{figure*}


\subsection{Results}
\label{sec:second_results}
The relation given by eq. \ref{eq:vg_NFW} allows us to map a given
$\vgs$ value onto the $\Mhs-\rss$ parameter space. This is shown in
Figure \ref{fig:vg_mh_rs}, upper panel. Note that for a given
  choice of the baryonic mass components of the potential, there is an
  absolute minimum for $\vgs$ (thereafter $V_{\rm G, min}$) , that
  corresponds to the absence of dark matter within 50 kpc. For our
  assumptions (eqs. \ref{eq_phi_b} and \ref{eq_phi_d}),
  $V_{\rm G, min} \approx 725$ \kms. In other words, this is the
  escape velocity from the GC only due to the deceleration imparted by
  the mass in the disc and bulge components.

  In Figure \ref{fig:vg_mh_rs}, the red dashed curve marks the
  iso-contour equal to $\vgs=850$ \kms: above this curve
  $V_{\rm G, min} \lsim \vgs < 850$ \kms. For a scale radius of
  $\rss < 30$ kpc, this region corresponds to
  $\Mhs < 1.5 \times 10^{12} M_{\odot}$, but, if larger \rs can be
  considered, the Milky Way mass can be larger. This parameter
  degeneracy is the result of fitting a measurement that --- as far as
  deceleration is concerned --- solely depends on the shape of the
  potential within 50 kpc: lighter, more concentrated haloes give the
  same net deceleration as more massive but less concentrated
  haloes. The $\vgs =850$ \kms line stands as an indicative limit
  above which, for a given halo mass, HVS data can be fitted at $>5\%$
  significance level assuming a B-type binary population in the GC
  close to that inferred in the LMC. In fact, since in our case
  $V_{\rm G, min} > 630$ \kms, the observed Galactic binary statistics
  never gives a high significance level fit to current data (see
  Section \ref{sec:1approach_results}).

To gain further insight into the likelihood of various regions of the
parameter space, we compare our results to additional Milky Way
observations and theoretical predictions. We compute the circular
velocity $V_{\rm c}=\sqrt{G M(<r)/r}$ along the Galactic disc plane,
where $M(<r)$ is the total enclosed mass (obtained integrating
eq.~\ref{eq:sumofpotentials}). We compare it to a recent compilation
of data from \citet{huang+16}, which traces the rotation curve of the
Milky Way out to $\sim 100$ kpc. Specifically, using a Markov Chain
Monte Carlo (MCMC) technique (see Appendix \ref{mcmc_appendix}), we
find that a relatively narrow region of the parameter space leads to a
fair description of the circular velocity data. As shown in the middle
panel of Fig. \ref{fig:vg_mh_rs}, the preferred combinations of \rs
and $\Mhs$ lie above our $\vgs \sim 850$ \kms iso-velocity line and
the best fitting parameters are
$\Mhs \approx 8 \times 10^{11} M_{\odot}$ and $\rss \approx 25$
kpc. More generally, \rs greater than $\sim 30$ ($\sim 35$) kpc for
our Galaxy can be excluded at, at least, one-sigma (two-sigma) level
(see also Figure \ref{fig:appendix} right panel).  This may be
intuitively understood as follows. At distances where dark matter
dominates, $\rss$ sets the scale beyond which
$V_{\rm c} \propto \sqrt{(M(<r)/r)} \sim \sqrt{\log{r}/r}$, while for
$r< \rss$ $V_{\rm c} \propto \sqrt{r}$. Therefore, a scale radius larger than $\sim 30$ kpc
cannot account for the observed rather flat/slowly decreasing
behaviour of the circular velocity at distances of $\gsim 20$ kpc (see
Figure \ref{fig:appendix} left panel). In addition, for a fixed
$M_{\rm h}$, large scale radii produce values of $V_{\rm c}$ lower
than the measured $V_{\rm c} \sim 200$ \kms in the halo region.

The lowest panel of Fig. \ref{fig:vg_mh_rs} shows the values of $\Mhs$
and $\rss$ found in the EAGLE hydro-cosmological simulation
\citep{schaye+15} and reported by \citet{schaller+15}.  The region of
parameter space within $\vgs < 850$ \kms and \rs $\lsim 35$ kpc fully
overlaps with the one-sigma and two-sigma regions determined using the
haloes in the EAGLE simulation. We also plot the $\Mhs$ and $\rss$
values that describe the halo in the Eris simulation \citep{guedes+11}
and note that they lie at the edge of the lowest two-sigma confidence
region.

\subsection{Impact of different disc and bulge models}
\label{sec:bulgedisc}
\begin{figure}
\includegraphics[width=0.5\textwidth]{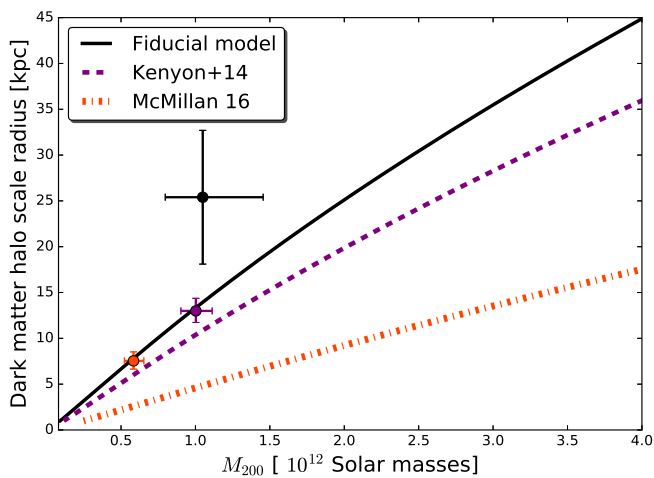}
\caption{Dark Halo mass ($M_{\rm 200}$) versus dark matter scale
    radius ($r_{\rm s}$) for 3 different models for the Galactic
    potential: the model presented in Section \ref{sec:NFW}
    (``Fiducial model"), the one adopted by \citet{kenyon+14} and one
    which combines our disc model and a symmetric average of the bulge
    matter density profile, as reported by \citet{mcmillan16}. The
    plotted lines are combinations of mass and radius that give an
    escape velocity from the GC of 850 \kms. Over-plotted in matching
    colours for each Galactic potential model are the best fitting
    parameters for the Galactic circular velocity (see Appendix
    \ref{mcmc_appendix}). Note that a mixed model with the
    \citet{mcmillan16}'s bulge and the Kenyon et al.'s parameters for
    the disc gives intermediate results.}
\label{fig:5}
\end{figure}
The mapping $\vgs \rightarrow (\Mhs-\rss)$ depends on the assumed
  baryonic matter density distribution, upon which there is no full
  general agreement \citep[see][for a recent observational review on
  the Galactic content and structure] {blad&gerhard16}. In particular,
  both the total baryonic mass and its concentration can have an
  impact.
  The most recent works point towards a stellar mass in the bulge
  around $ 1-2 \times 10^{10} M_{\odot}$ \citep[e.g.][]{portail+15},
  but one should be aware of uncertainties given by the fact that
  different observational studies of the bulge constrain the mass in
  different regions and the size of the bulge is not universally
  defined. Moreover, the bulge's mass is distributed in a complex
  box/peanut structure, coexisting with an addition spherical
  component \citep[see][for an observational review on the
  bulge]{gonzalez&gadotti16}. The corresponding 3-dimensional density
  profile down to the sphere of influence of Sgr A*, is therefore
  uncertain.  Likewise for the disc component, there are ongoing
  efforts to try and construct a fully consistent picture, that is
  currently missing \citep[see][for a recent review on the stellar
  disc]{rix&bovy13}.  Recent estimates place the total disc mass
  around $ 5 \times 10^{10} M_{\odot}$, a factor of two lighter than
  the disc mass we adopt in Fig.\ref{fig:vg_mh_rs}.

  Given these uncertainties, we here explore the impact of adopting
  different baryonic components than the ones we assumed in Section
  \ref{sec:NFW}, where a justification for that choices is stated. In
  particular, we explore lighter components, differently distributed.
  To do this, we compare in Figure \ref{fig:5} the loci of
  $\vgs = 850$ \kms in the plane $(M_{\rm 200}-\rss)$, given by other
  two Galactic potential models that together with ours should frame a
  plausible uncertainty range. We chose to plot here
  $M_{\rm 200}$\footnote{This is the mass enclosed within a sphere of
    mean density equal to 200 times the critical density of the
    Universe at $z=0$} instead of $M_{\rm h}$ as it is commonly used
  to indicate the Milky Way dark matter mass and it can facilitate
  comparisons with results from other probes.

  The potential adopted by \cite{kenyon+14} and widely used in the HVS
  community is shown with a dashed line: the bulge and disc components
  are described by our eqs. \ref{eq_phi_b} and \ref{eq_phi_d} but with
  different parameters ($M_{\rm b} = 3.76 \times 10^{9} M_{\odot}$,
  $r_{\rm b} = 0.1$ kpc, $M_{\rm d} = 6 \times 10^{10} M_{\odot}$,
  $a = 2.75$ kpc, $b = 0.3$ kpc).  Comparing the solid and dashed
  lines one concludes that, for a given $\rss$, the Kenyon et al.'s
  model gives $\sim 30\%$ more massive haloes.  We then calculate the
  $\vgs = 850$ \kms iso-courve for a bulge potential advocated by
  \citet{mcmillan16} plus our fiducial model for the disc (dash-dotted
  line).  The McMillan's bulge model adopts a total mass of
  $\simeq 8.9 \times 10^{9} M_{\odot}$ and it is not spherically
  symmetric. We therefore radially average the axisymmetric density
  profile before computing the corresponding
  potential\footnote{Indeed, we are comparing our models with a
    radially averaged observed distribution of HVS velocities beyond
    50 kpc, we can therefore assume a spherically symmetric bulge,
    since its spatial extension is no more than a few kpc.}. Note that
  the McMillan's bulge model is more massive than the Kenyon et al.'s
  one but equally concentrated, resulting in a very different density
  profile.  Consequently, this model gives significantly more massive
  haloes (by a factor $\gsim 2$) than we obtain with either Kenyon et
  al.'s or our fiducial model.

  We conclude that the impact of these uncertainties on the
  determination of the halo mass with HVS data is large and cannot be
  ignored.  In order to put robust constraints on the dark matter halo
  of our Galaxy through our method a multi-parameter fit of data is
  therefore required where both the disc and bulge parameters need to
  be left free to vary. We defer this more sophisticated analyses, however, when 
   more and better HVS data will be available.

  On the positive side, the main features of the two regions in the
  $\Mhs-\rss$ parameter space defined by our $\vgs = 850$ \kms remain
  the same, regardless of the specific baryonic potentials: the best
  fitting models for the circular velocity data always lie within the
  $\vgs < 850$ \kms region (see crosses in Figure \ref{fig:5} and
  Appendix \ref{mcmc_appendix}), as do the EAGLE's predictions for
  $\Lambda$CDM compatible haloes.
\begin{figure*}
\centering
  \includegraphics[width=\textwidth]{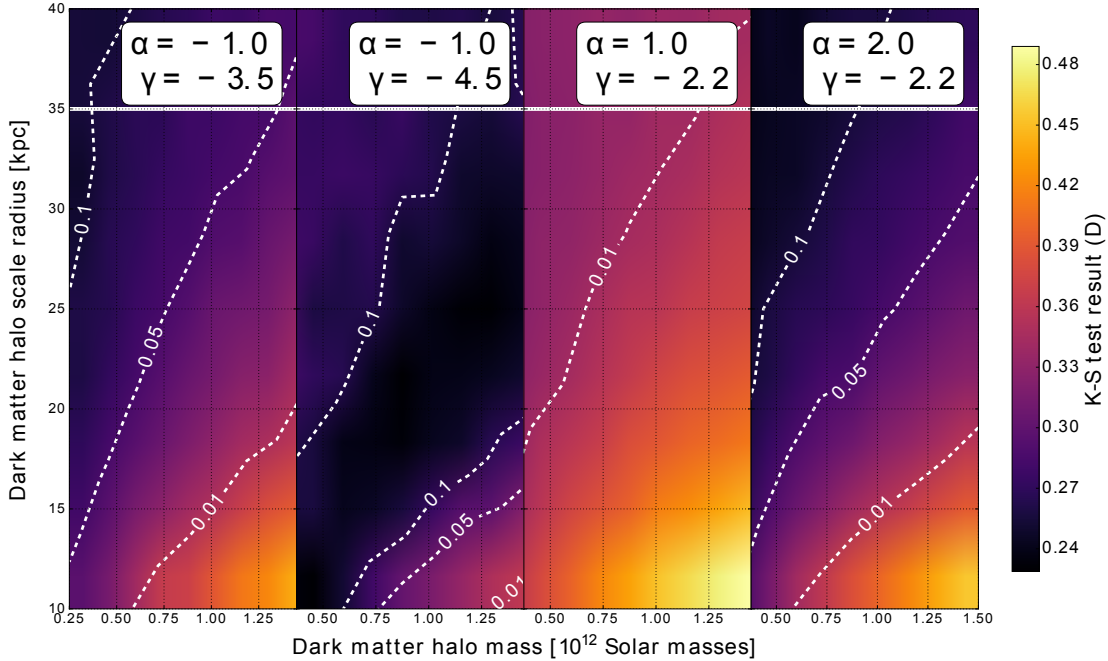}
  \caption{Contour plots for K-S test results in the parameter space
    $\Mhs-\rss$, for fixed $\alpha,\gamma$ pairs (see panels' label
    and star marks in Figure \ref{fig:vg_mh_rs}). Velocity
    distributions are computed radially decelerating each star in a
    given potential (see Section \ref{sec:NFW}). The white dashed
    lines are iso-contour lines for a given significance level
    $\bar{\alpha}$. Regions at the left of
    of each line have a value of $\bar{\alpha}$ larger than that
    stated in the corresponding label.}
\label{fig:Mh_rs_KS}
\end{figure*}


\section{Discussion and conclusions} \label{discussion}

The analysis presented in the paper yields the following main results:

\begin{itemize}
\item[1] For a $> 5\%$ ($>1\%$) significance level fit, HVS velocity
  data {\em alone} require a Galactic potential with an escape
  velocity from the GC to 50 kpc $\lsim 850$ \kms ($\lsim 930$ \kms),
  when assuming that binary stars within the innermost few parsecs of
  our Galaxy are not dissimilar from binaries in other, more
  observationally accessible {\it star forming} regions. For
  $\vgs \sim 630$ \kms, the binary statistics for late B-type stars
  observed in the Solar neighbourhood also provide a fit at the same
  significance level.

\item[2] When specialising to a NFW dark matter halo, we find that the
  region $V_{\rm G} \lsim 850$ \kms contains models that are
  compatible with both HVS and circular velocity data. These models
  also correspond to $\Lambda$CDM-compatible Milky Way haloes. In
  principle, we cannot exclude the parameter space $\vgs \gsim 850$
  \kms. However, it would require us to face both an increasingly
  different statistical description of the binary population in the GC
  with respect to current observations {\em and} dark matter haloes
  that are inconsistent with predictions in the $\Lambda$CDM model at
  one-sigma level or more (see lower panel of Figure
  \ref{fig:vg_mh_rs}).

\item[3] The result stated in point 2 is {\it independent} of the assumed
   baryonic components of the Galactic potential,
  across a wide range for plausible masses and scale radii.

\item[4] However, the {\it specific} mapping of $\vgs$ values onto the
  $\Mhs-\rss$ parameter space {\it is} highly dependent on the assumed
  bulge and disc models (see Section \ref{sec:bulgedisc}).  Both the
  baryonic total mass and its distribution affect the results.  In
  general, works that try to infer the dark matter halo mass from HVS
  data should fold in the uncertainties linked to our imperfect
  knowledge of the baryonic mass distribution.
\end{itemize}

These results rely on certain assumptions for the binary population in
the GC whose impact we now discuss. Following the same computational
procedure previously presented for our fiducial model, we have found
that a different mass function for the primary stars (either a
Salpeter or a top-heavy mass function) or a change in metallicity
(from super-solar to solar) do not substantially alter our
results. However, the choice of the minimum companion mass
(i.e. $m_{\rm min}$ in eq. \ref{eq:fq}) does lead to different
conclusions. In particular, the higher $m_{\rm min}$, the steeper the
binary distributions should be to fit the data, even for low ($< 850$
\kms) $\vgs$. For example, for $m_{\rm min} = 0.3 M_{\odot}$
  (instead of 0.1 $M_{\odot}$) and $\vgs = 760$ \kms the stripe of
  minima for the K-S test runs along the $\gamma \approx - 6.5$ and
  $\alpha \approx 4.5$ directions, very far from the observed values.
Currently, there is no observational or theoretical reason why we
should adopt a higher minimum mass than the one usually assumed (``the
brown dwarf'' limit), but this exercise shows that better quality and
quantity HVS data has the potential to statistically constrain the
minimum mass for a secondary, which may shed light on star and/or
binary forming mechanisms at work in the GC.

A second set of uncertainties that may affect our conclusions pertain
to the observed binary parameter distributions in the 30 Doradus
region, that we use as guidance. The 30 Doradus B-type sample of
\citet{dunstall+15} is based on 6 epochs of spectra, that do not allow
for a full orbital solution for each system. These authors' results
are mainly based on the distribution of the maximum variation in
radial velocities per system, from where they statistically derive
constraints for the full sample. Another point worth stressing is that
the 30 Doradus B-type sample is of {\em early} type stars (mass roughly around
$10 M_{\odot}$) and distributions for {\em late} B-type star binaries
in star forming regions may be different. However, these latter are
not currently available, and therefore the Dustall et al. sample
remains the most relevant to guide our analysis in those
regions. Our statement is therefore that the statistical distributions
derived from this sample (including the statistical errors on the
power-law indexes) can reproduce HVS data at a several percentage
confidence level. Far more reliable is the statistical
  description of observed late B-type binaries in the Solar
  neighbourhood, that can be easily reconciled with HVS data {\it
    only} for quite low $\vgs$ potentials.

A possibility that we have not so far discussed is that dynamical
processes that inject binaries within Sgr A*'s tidal sphere modify the
natal mass ratio and separation distributions. Unfortunately, as far
as we know, dedicated studies are missing and we will then only
discuss the consequence of the classical loss-cone\footnote{The loss
  cone theory deals with processes by which stars are ``lost'' because
  they enter the tidal sphere, in which they will suffer tidal
  disruption on a dynamical time. The name comes from the fact that
  the tidal sphere is defined in velocity space at a fixed position as
  a ``cone" with an angle proportional to the angular momentum needed
  for the (binary) star to be put on an orbit grazing the tidal radius
  \cite[see for e.g.][section 6.1.1]{alexander05}.} theory'' dealing
with two-body encounters \cite[e.g.][]{F&R76,L&S77} as derived in
\citet[][section 3]{rossi+14}. Their considerations show that even
allowing for extreme regimes, one would expect no modification in the
mass ratio distribution and a modification in the separation
distribution by no more than a factor of ``a'' (i.e. a natal \"Opik's
law would evolve into $f_{\rm a} \sim$ const.). This would 
  increase the $\vgs$ range ($\vgs \lsim 750$ \kms) compatible with
  Solar neighbourhood observations (see Fig.~\ref{fig:vGs}). Beside
  that, all our results remain unchanged.

We would also like to remark here that, although observed binary
  parameters give acceptable fits for $\vgs< 930$ \kms, the K-S test
results currently prefer even steeper mass ratio and binary
separation distributions ($\gamma \sim -4.5$ instead of
$\gamma \sim-3.5$ and/or $\alpha \sim 2$ instead of -1, see
Fig. \ref{fig:Mh_rs_KS}). This larger $|\gamma|$ value gives a steeper
high velocity tail, which better match the lack of observed $> 700$
\kms HVSs. From the above considerations, modification of the natal
distribution by standard two-body scattering into the binary loss cone
may not be held responsible. Assuming that the halo actually has
$\vgs < 930$ \kms, one possible inference is indeed that
$\gamma \sim -4.5$ is a better description of the B-type binary natal
distribution in the GC, close but not identical to that in the
Tarantula Nebula.

It is of course possible that some other dynamical interactions
(e.g. binary softening/hardening, collisions) or disruption of
binaries in triples could be indeed responsible for a change in
$\gamma$ and a larger one in $\alpha$.  However, for massive binaries
dynamical evolution of their properties may be neglected in the GC,
because it would happen on timescales longer than their lifetime
\citep{pfuhl+14}. On the contrary, it may be relevant for low mass
binaries, but only within the inner 0.1 pc
\citep{hopman09}. Nevertheless, these possibilities would be very
intriguing to explore in depth, if more and better data on HVSs
together with a more solid knowledge of binary properties in different
regions will still indicate the need for such processes.

Finally, given the paucity of data, we did not use any spatial
distribution information but we rather fitted the velocity
distribution integrated over the observed radial range. This precluded
the possibility to meaningfully investigate anisotropic dark matter
distributions and we preferred to confine ourselves to spherically
symmetric potentials.

All the above uncertainties and possibilities can and should be tested
and explored when a HVS data sample that extends below and above the
velocity peak is available. Such a data set would allow us to break
the degeneracy between halo and binary parameters, as the rise to the
peak and the peak itself are mostly sensitive to the halo properties,
whereas the high velocity tail is primarily shaped by the binary
distributions. This will be achieved in the coming few years thanks to
the ESA mission Gaia, whose catalogue should contain at least a few
hundred HVSs with precise astrometric measurements. Moreover Gaia
will greatly improve our knowledge of binary statistics in the Galaxy
(but not directly in the GC, where infrared observations are required)
and in the LMC allowing us to draw more robust inferences. 

In conclusion, this paper shows for the first time the potential of
HVS data combined with our modelling method to extract joint
information on the GC and (dark) matter distribution. It is clear,
however, that the full realisation of this potential requires a larger
and less biased set of data. The ESA Gaia mission is likely to provide
such a sample within the coming five years.


\section*{Acknowledgements} We thank ~S. de Mink, R. Sch\"odel
  for a careful reading of the manuscript and O. Gnedin, M. Viola,
  H. Perets, E. Starkenburg, J. Navarro, A. Helmi, S. Kobayashi and
  A. Brown for useful discussions and comments.  We also thank the
  anonymous referee for the careful reading of the manuscript and
  his/her useful suggestions.
E.M.R. and T.M acknowledge the supported from NWO TOP grant Module 2, project number  614.001.401. 


\bibliographystyle{mnras}
\bibliography{hvs_halo}

\begin{thebibliography}{}
\makeatletter
\relax
\def\mn@urlcharsother{\let\do\@makeother \do\$\do\&\do\#\do\^\do\_\do\%\do\~}
\def\mn@doi{\begingroup\mn@urlcharsother \@ifnextchar [ {\mn@doi@}
  {\mn@doi@[]}}
\def\mn@doi@[#1]#2{\def\@tempa{#1}\ifx\@tempa\@empty \href
  {http://dx.doi.org/#2} {doi:#2}\else \href {http://dx.doi.org/#2} {#1}\fi
  \endgroup}
\def\mn@eprint#1#2{\mn@eprint@#1:#2::\@nil}
\def\mn@eprint@arXiv#1{\href {http://arxiv.org/abs/#1} {{\tt arXiv:#1}}}
\def\mn@eprint@dblp#1{\href {http://dblp.uni-trier.de/rec/bibtex/#1.xml}
  {dblp:#1}}
\def\mn@eprint@#1:#2:#3:#4\@nil{\def\@tempa {#1}\def\@tempb {#2}\def\@tempc
  {#3}\ifx \@tempc \@empty \let \@tempc \@tempb \let \@tempb \@tempa \fi \ifx
  \@tempb \@empty \def\@tempb {arXiv}\fi \@ifundefined
  {mn@eprint@\@tempb}{\@tempb:\@tempc}{\expandafter \expandafter \csname
  mn@eprint@\@tempb\endcsname \expandafter{\@tempc}}}

\bibitem[\protect\citeauthoryear{{Aarseth}}{{Aarseth}}{1974}]{aarseth74}
{Aarseth} S.~J.,  1974, \aap, \href
  {http://adsabs.harvard.edu/abs/1974A%26A....35..237A} {35, 237}

\bibitem[\protect\citeauthoryear{{Akeret}, {Seehars}, {Amara}, {Refregier}  \&
  {Csillaghy}}{{Akeret} et~al.}{2013}]{Akaret13}
{Akeret} J.,  {Seehars} S.,  {Amara} A.,  {Refregier} A.,   {Csillaghy} A.,
  2013, \mn@doi [Astronomy and Computing] {10.1016/j.ascom.2013.06.003}, \href
  {http://adsabs.harvard.edu/abs/2013A%26C.....2...27A} {2, 27}

\bibitem[\protect\citeauthoryear{{Alexander}}{{Alexander}}{2005}]{alexander05}
{Alexander} T.,  2005, \mn@doi [\physrep] {10.1016/j.physrep.2005.08.002},
  \href {http://adsabs.harvard.edu/abs/2005PhR...419...65A} {419, 65}

\bibitem[\protect\citeauthoryear{{Bartko} et~al.,}{{Bartko}
  et~al.}{2010}]{bartko+10}
{Bartko} H.,  et~al., 2010, \mn@doi [\apj] {10.1088/0004-637X/708/1/834}, \href
  {http://adsabs.harvard.edu/abs/2010ApJ...708..834B} {708, 834}

\bibitem[\protect\citeauthoryear{{Blaauw}}{{Blaauw}}{1961}]{blaauw61}
{Blaauw} A.,  1961, \bain, \href
  {http://adsabs.harvard.edu/abs/1961BAN....15..265B} {15, 265}

\bibitem[\protect\citeauthoryear{{Bland-Hawthorn} \&
  {Gerhard}}{{Bland-Hawthorn} \& {Gerhard}}{2016}]{blad&gerhard16}
{Bland-Hawthorn} J.,  {Gerhard} O.,  2016, preprint, \href
  {http://adsabs.harvard.edu/abs/2016arXiv160207702B} {} (\mn@eprint {arXiv}
  {1602.07702})

\bibitem[\protect\citeauthoryear{{B{\"o}ker}}{{B{\"o}ker}}{2010}]{boeker10}
{B{\"o}ker} T.,  2010, in {de Grijs} R.,  {L{\'e}pine} J.~R.~D.,  eds,  IAU
  Symposium Vol. 266, Star Clusters: Basic Galactic Building Blocks Throughout
  Time and Space. pp 58--63 (\mn@eprint {arXiv} {0910.4863}),
  \mn@doi{10.1017/S1743921309990871}

\bibitem[\protect\citeauthoryear{{Boylan-Kolchin}, {Bullock}  \&
  {Kaplinghat}}{{Boylan-Kolchin} et~al.}{2011}]{boylan+11}
{Boylan-Kolchin} M.,  {Bullock} J.~S.,   {Kaplinghat} M.,  2011, \mn@doi
  [\mnras] {10.1111/j.1745-3933.2011.01074.x}, \href
  {http://adsabs.harvard.edu/abs/2011MNRAS.415L..40B} {415, L40}

\bibitem[\protect\citeauthoryear{{Bromley}, {Kenyon}, {Brown}  \&
  {Geller}}{{Bromley} et~al.}{2009}]{bromley+09}
{Bromley} B.~C.,  {Kenyon} S.~J.,  {Brown} W.~R.,   {Geller} M.~J.,  2009,
  \mn@doi [\apj] {10.1088/0004-637X/706/2/925}, \href
  {http://adsabs.harvard.edu/abs/2009ApJ...706..925B} {706, 925}

\bibitem[\protect\citeauthoryear{{Brown}}{{Brown}}{2015}]{brown15}
{Brown} W.~R.,  2015, \mn@doi [\araa] {10.1146/annurev-astro-082214-122230},
  \href {http://adsabs.harvard.edu/abs/2015ARA%26A..53...15B} {53, 15}

\bibitem[\protect\citeauthoryear{{Brown}, {Geller}  \& {Kenyon}}{{Brown}
  et~al.}{2014}]{brown+14}
{Brown} W.~R.,  {Geller} M.~J.,   {Kenyon} S.~J.,  2014, \mn@doi [\apj]
  {10.1088/0004-637X/787/1/89}, \href
  {http://adsabs.harvard.edu/abs/2014ApJ...787...89B} {787, 89}

\bibitem[\protect\citeauthoryear{{Bullock}, {Stewart}, {Kaplinghat}, {Tollerud}
   \& {Wolf}}{{Bullock} et~al.}{2010}]{bullock+10}
{Bullock} J.~S.,  {Stewart} K.~R.,  {Kaplinghat} M.,  {Tollerud} E.~J.,
  {Wolf} J.,  2010, \mn@doi [\apj] {10.1088/0004-637X/717/2/1043}, \href
  {http://adsabs.harvard.edu/abs/2010ApJ...717.1043B} {717, 1043}

\bibitem[\protect\citeauthoryear{{Carollo}, {Stiavelli}, {de Zeeuw}  \&
  {Mack}}{{Carollo} et~al.}{1997}]{carollo+97}
{Carollo} C.~M.,  {Stiavelli} M.,  {de Zeeuw} P.~T.,   {Mack} J.,  1997,
  \mn@doi [\aj] {10.1086/118654}, \href
  {http://adsabs.harvard.edu/abs/1997AJ....114.2366C} {114, 2366}

\bibitem[\protect\citeauthoryear{{Carson}, {Barth}, {Seth}, {den Brok},
  {Cappellari}, {Greene}, {Ho}  \& {Neumayer}}{{Carson}
  et~al.}{2015}]{carson+15}
{Carson} D.~J.,  {Barth} A.~J.,  {Seth} A.~C.,  {den Brok} M.,  {Cappellari}
  M.,  {Greene} J.~E.,  {Ho} L.~C.,   {Neumayer} N.,  2015, \mn@doi [\aj]
  {10.1088/0004-6256/149/5/170}, \href
  {http://adsabs.harvard.edu/abs/2015AJ....149..170C} {149, 170}

\bibitem[\protect\citeauthoryear{{Do}, {Kerzendorf}, {Winsor}, {St{\o}stad},
  {Morris}, {Lu}  \& {Ghez}}{{Do} et~al.}{2015}]{do+15}
{Do} T.,  {Kerzendorf} W.,  {Winsor} N.,  {St{\o}stad} M.,  {Morris} M.~R.,
  {Lu} J.~R.,   {Ghez} A.~M.,  2015, \mn@doi [\apj]
  {10.1088/0004-637X/809/2/143}, \href
  {http://adsabs.harvard.edu/abs/2015ApJ...809..143D} {809, 143}

\bibitem[\protect\citeauthoryear{{Duch{\^e}ne} \& {Kraus}}{{Duch{\^e}ne} \&
  {Kraus}}{2013}]{D&K13}
{Duch{\^e}ne} G.,  {Kraus} A.,  2013, \mn@doi [\araa]
  {10.1146/annurev-astro-081710-102602}, \href
  {http://adsabs.harvard.edu/abs/2013ARA%26A..51..269D} {51, 269}

\bibitem[\protect\citeauthoryear{{Dunstall} et~al.,}{{Dunstall}
  et~al.}{2015}]{dunstall+15}
{Dunstall} P.~R.,  et~al., 2015, \mn@doi [\aap] {10.1051/0004-6361/201526192},
  \href {http://adsabs.harvard.edu/abs/2015A%26A...580A..93D} {580, A93}

\bibitem[\protect\citeauthoryear{{Eldridge}, {Langer}  \& {Tout}}{{Eldridge}
  et~al.}{2011}]{eldridge+11}
{Eldridge} J.~J.,  {Langer} N.,   {Tout} C.~A.,  2011, \mn@doi [\mnras]
  {10.1111/j.1365-2966.2011.18650.x}, \href
  {http://adsabs.harvard.edu/abs/2011MNRAS.414.3501E} {414, 3501}

\bibitem[\protect\citeauthoryear{{Foreman-Mackey}, {Hogg}, {Lang}  \&
  {Goodman}}{{Foreman-Mackey} et~al.}{2013}]{Foreman13}
{Foreman-Mackey} D.,  {Hogg} D.~W.,  {Lang} D.,   {Goodman} J.,  2013, \mn@doi
  [\pasp] {10.1086/670067}, \href
  {http://adsabs.harvard.edu/abs/2013PASP..125..306F} {125, 306}

\bibitem[\protect\citeauthoryear{{Fragione} \& {Loeb}}{{Fragione} \&
  {Loeb}}{2016}]{fragione&loeb16}
{Fragione} G.,  {Loeb} A.,  2016, preprint, \href
  {http://adsabs.harvard.edu/abs/2016arXiv160801517F} {} (\mn@eprint {arXiv}
  {1608.01517})

\bibitem[\protect\citeauthoryear{{Frank} \& {Rees}}{{Frank} \&
  {Rees}}{1976}]{F&R76}
{Frank} J.,  {Rees} M.~J.,  1976, \mn@doi [\mnras] {10.1093/mnras/176.3.633},
  \href {http://adsabs.harvard.edu/abs/1976MNRAS.176..633F} {176, 633}

\bibitem[\protect\citeauthoryear{{Fritz} et~al.,}{{Fritz}
  et~al.}{2016}]{fritz+16}
{Fritz} T.~K.,  et~al., 2016, \mn@doi [\apj] {10.3847/0004-637X/821/1/44},
  \href {http://adsabs.harvard.edu/abs/2016ApJ...821...44F} {821, 44}

\bibitem[\protect\citeauthoryear{{Genzel}, {Eisenhauer}  \&
  {Gillessen}}{{Genzel} et~al.}{2010}]{genzel+10}
{Genzel} R.,  {Eisenhauer} F.,   {Gillessen} S.,  2010, \mn@doi [Reviews of
  Modern Physics] {10.1103/RevModPhys.82.3121}, \href
  {http://adsabs.harvard.edu/abs/2010RvMP...82.3121G} {82, 3121}

\bibitem[\protect\citeauthoryear{{Georgiev} \& {B{\"o}ker}}{{Georgiev} \&
  {B{\"o}ker}}{2014}]{georvievboeker14}
{Georgiev} I.~Y.,  {B{\"o}ker} T.,  2014, \mn@doi [\mnras]
  {10.1093/mnras/stu797}, \href
  {http://adsabs.harvard.edu/abs/2014MNRAS.441.3570G} {441, 3570}

\bibitem[\protect\citeauthoryear{{Ghez} et~al.,}{{Ghez} et~al.}{2008}]{ghez+08}
{Ghez} A.~M.,  et~al., 2008, \mn@doi [\apj] {10.1086/592738}, \href
  {http://adsabs.harvard.edu/abs/2008ApJ...689.1044G} {689, 1044}

\bibitem[\protect\citeauthoryear{{Gibbons}, {Belokurov}  \& {Evans}}{{Gibbons}
  et~al.}{2014}]{gibbons+14}
{Gibbons} S.~L.~J.,  {Belokurov} V.,   {Evans} N.~W.,  2014, \mn@doi [\mnras]
  {10.1093/mnras/stu1986}, \href
  {http://adsabs.harvard.edu/abs/2014MNRAS.445.3788G} {445, 3788}

\bibitem[\protect\citeauthoryear{{Gillessen}, {Eisenhauer}, {Trippe},
  {Alexander}, {Genzel}, {Martins}  \& {Ott}}{{Gillessen}
  et~al.}{2009}]{gillessen+09}
{Gillessen} S.,  {Eisenhauer} F.,  {Trippe} S.,  {Alexander} T.,  {Genzel} R.,
  {Martins} F.,   {Ott} T.,  2009, \mn@doi [\apj]
  {10.1088/0004-637X/692/2/1075}, \href
  {http://adsabs.harvard.edu/abs/2009ApJ...692.1075G} {692, 1075}

\bibitem[\protect\citeauthoryear{{Gnedin}, {Gould}, {Miralda-Escud{\'e}}  \&
  {Zentner}}{{Gnedin} et~al.}{2005}]{gnedin+05}
{Gnedin} O.~Y.,  {Gould} A.,  {Miralda-Escud{\'e}} J.,   {Zentner} A.~R.,
  2005, \mn@doi [\apj] {10.1086/496958}, \href
  {http://adsabs.harvard.edu/abs/2005ApJ...634..344G} {634, 344}

\bibitem[\protect\citeauthoryear{{Gnedin}, {Brown}, {Geller}  \&
  {Kenyon}}{{Gnedin} et~al.}{2010}]{gnedin+10}
{Gnedin} O.~Y.,  {Brown} W.~R.,  {Geller} M.~J.,   {Kenyon} S.~J.,  2010,
  \mn@doi [\apjl] {10.1088/2041-8205/720/1/L108}, \href
  {http://adsabs.harvard.edu/abs/2010ApJ...720L.108G} {720, L108}

\bibitem[\protect\citeauthoryear{{Gonzalez} \& {Gadotti}}{{Gonzalez} \&
  {Gadotti}}{2016}]{gonzalez&gadotti16}
{Gonzalez} O.~A.,  {Gadotti} D.,  2016, \mn@doi [Galactic Bulges]
  {10.1007/978-3-319-19378-6_9}, \href
  {http://adsabs.harvard.edu/abs/2016ASSL..418..199G} {418, 199}

\bibitem[\protect\citeauthoryear{{Goodman} \& {Weare}}{{Goodman} \&
  {Weare}}{2010}]{Goodman10}
{Goodman} J.,  {Weare} J.,  2010, \mn@doi [Comm. App. Math. Comp. Sci.]
  {10.2140/camcos.2010.5.65}, 5, 65

\bibitem[\protect\citeauthoryear{{Guedes}, {Callegari}, {Madau}  \&
  {Mayer}}{{Guedes} et~al.}{2011}]{guedes+11}
{Guedes} J.,  {Callegari} S.,  {Madau} P.,   {Mayer} L.,  2011, \mn@doi [\apj]
  {10.1088/0004-637X/742/2/76}, \href
  {http://adsabs.harvard.edu/abs/2011ApJ...742...76G} {742, 76}

\bibitem[\protect\citeauthoryear{{Hawkins} et~al.,}{{Hawkins}
  et~al.}{2015}]{hawkins+15}
{Hawkins} K.,  et~al., 2015, \mn@doi [\mnras] {10.1093/mnras/stu2574}, \href
  {http://adsabs.harvard.edu/abs/2015MNRAS.447.2046H} {447, 2046}

\bibitem[\protect\citeauthoryear{{Hernquist}}{{Hernquist}}{1990}]{hernquist90}
{Hernquist} L.,  1990, \mn@doi [\apj] {10.1086/168845}, \href
  {http://adsabs.harvard.edu/abs/1990ApJ...356..359H} {356, 359}

\bibitem[\protect\citeauthoryear{{Hills}}{{Hills}}{1988}]{hills88}
{Hills} J.~G.,  1988, \mn@doi [\nat] {10.1038/331687a0}, \href
  {http://adsabs.harvard.edu/abs/1988Natur.331..687H} {331, 687}

\bibitem[\protect\citeauthoryear{{Hopman}}{{Hopman}}{2009}]{hopman09}
{Hopman} C.,  2009, \mn@doi [\apj] {10.1088/0004-637X/700/2/1933}, \href
  {http://adsabs.harvard.edu/abs/2009ApJ...700.1933H} {700, 1933}

\bibitem[\protect\citeauthoryear{{Huang} et~al.,}{{Huang}
  et~al.}{2016}]{huang+16}
{Huang} Y.,  et~al., 2016, preprint, \href
  {http://adsabs.harvard.edu/abs/2016arXiv160401216H} {} (\mn@eprint {arXiv}
  {1604.01216})

\bibitem[\protect\citeauthoryear{{Hurley}, {Pols}  \& {Tout}}{{Hurley}
  et~al.}{2000}]{hurley+00}
{Hurley} J.~R.,  {Pols} O.~R.,   {Tout} C.~A.,  2000, \mn@doi [\mnras]
  {10.1046/j.1365-8711.2000.03426.x}, \href
  {http://adsabs.harvard.edu/abs/2000MNRAS.315..543H} {315, 543}

\bibitem[\protect\citeauthoryear{{Johnston}, {Spergel}  \&
  {Hernquist}}{{Johnston} et~al.}{1995}]{johnston+95}
{Johnston} K.~V.,  {Spergel} D.~N.,   {Hernquist} L.,  1995, \mn@doi [\apj]
  {10.1086/176247}, \href {http://adsabs.harvard.edu/abs/1995ApJ...451..598J}
  {451, 598}

\bibitem[\protect\citeauthoryear{{Kafle}, {Sharma}, {Lewis}  \&
  {Bland-Hawthorn}}{{Kafle} et~al.}{2014}]{kafle+14}
{Kafle} P.~R.,  {Sharma} S.,  {Lewis} G.~F.,   {Bland-Hawthorn} J.,  2014,
  \mn@doi [\apj] {10.1088/0004-637X/794/1/59}, \href
  {http://adsabs.harvard.edu/abs/2014ApJ...794...59K} {794, 59}

\bibitem[\protect\citeauthoryear{{Kenyon}, {Bromley}, {Geller}  \&
  {Brown}}{{Kenyon} et~al.}{2008}]{kenyon+08}
{Kenyon} S.~J.,  {Bromley} B.~C.,  {Geller} M.~J.,   {Brown} W.~R.,  2008,
  \mn@doi [\apj] {10.1086/587738}, \href
  {http://adsabs.harvard.edu/abs/2008ApJ...680..312K} {680, 312}

\bibitem[\protect\citeauthoryear{{Kenyon}, {Bromley}, {Brown}  \&
  {Geller}}{{Kenyon} et~al.}{2014}]{kenyon+14}
{Kenyon} S.~J.,  {Bromley} B.~C.,  {Brown} W.~R.,   {Geller} M.~J.,  2014,
  \mn@doi [\apj] {10.1088/0004-637X/793/2/122}, \href
  {http://adsabs.harvard.edu/abs/2014ApJ...793..122K} {793, 122}

\bibitem[\protect\citeauthoryear{{Klypin}, {Kravtsov}, {Valenzuela}  \&
  {Prada}}{{Klypin} et~al.}{1999}]{klypin+99}
{Klypin} A.,  {Kravtsov} A.~V.,  {Valenzuela} O.,   {Prada} F.,  1999, \mn@doi
  [\apj] {10.1086/307643}, \href
  {http://adsabs.harvard.edu/abs/1999ApJ...522...82K} {522, 82}

\bibitem[\protect\citeauthoryear{{Kobayashi}, {Hainick}, {Sari}  \&
  {Rossi}}{{Kobayashi} et~al.}{2012}]{kobayashi+12}
{Kobayashi} S.,  {Hainick} Y.,  {Sari} R.,   {Rossi} E.~M.,  2012, \mn@doi
  [\apj] {10.1088/0004-637X/748/2/105}, \href
  {http://adsabs.harvard.edu/abs/2012ApJ...748..105K} {748, 105}

\bibitem[\protect\citeauthoryear{{Kobulnicky} et~al.,}{{Kobulnicky}
  et~al.}{2014}]{kobulnicky+14}
{Kobulnicky} H.~A.,  et~al., 2014, \mn@doi [\apjs]
  {10.1088/0067-0049/213/2/34}, \href
  {http://adsabs.harvard.edu/abs/2014ApJS..213...34K} {213, 34}

\bibitem[\protect\citeauthoryear{{Kouwenhoven}, {Brown}, {Portegies Zwart}  \&
  {Kaper}}{{Kouwenhoven} et~al.}{2007}]{kouwenhoven+07}
{Kouwenhoven} M.~B.~N.,  {Brown} A.~G.~A.,  {Portegies Zwart} S.~F.,   {Kaper}
  L.,  2007, \mn@doi [\aap] {10.1051/0004-6361:20077719}, \href
  {http://adsabs.harvard.edu/abs/2007A%26A...474...77K} {474, 77}

\bibitem[\protect\citeauthoryear{{Kroupa}}{{Kroupa}}{2002}]{kroupa02}
{Kroupa} P.,  2002, \mn@doi [Science] {10.1126/science.1067524}, \href
  {http://adsabs.harvard.edu/abs/2002Sci...295...82K} {295, 82}

\bibitem[\protect\citeauthoryear{{Laevens} et~al.,}{{Laevens}
  et~al.}{2015}]{laevens+15}
{Laevens} B.~P.~M.,  et~al., 2015, \mn@doi [\apj] {10.1088/0004-637X/813/1/44},
  \href {http://adsabs.harvard.edu/abs/2015ApJ...813...44L} {813, 44}

\bibitem[\protect\citeauthoryear{{Law} \& {Majewski}}{{Law} \&
  {Majewski}}{2010}]{L&M10}
{Law} D.~R.,  {Majewski} S.~R.,  2010, \mn@doi [\apj]
  {10.1088/0004-637X/714/1/229}, \href
  {http://adsabs.harvard.edu/abs/2010ApJ...714..229L} {714, 229}

\bibitem[\protect\citeauthoryear{{Lightman} \& {Shapiro}}{{Lightman} \&
  {Shapiro}}{1977}]{L&S77}
{Lightman} A.~P.,  {Shapiro} S.~L.,  1977, \mn@doi [\apj] {10.1086/154925},
  \href {http://adsabs.harvard.edu/abs/1977ApJ...211..244L} {211, 244}

\bibitem[\protect\citeauthoryear{{Loebman} et~al.,}{{Loebman}
  et~al.}{2014}]{loebman+14}
{Loebman} S.~R.,  et~al., 2014, \mn@doi [\apj] {10.1088/0004-637X/794/2/151},
  \href {http://adsabs.harvard.edu/abs/2014ApJ...794..151L} {794, 151}

\bibitem[\protect\citeauthoryear{{Lu}, {Do}, {Ghez}, {Morris}, {Yelda}  \&
  {Matthews}}{{Lu} et~al.}{2013}]{lu+13}
{Lu} J.~R.,  {Do} T.,  {Ghez} A.~M.,  {Morris} M.~R.,  {Yelda} S.,   {Matthews}
  K.,  2013, \mn@doi [\apj] {10.1088/0004-637X/764/2/155}, \href
  {http://adsabs.harvard.edu/abs/2013ApJ...764..155L} {764, 155}

\bibitem[\protect\citeauthoryear{{Madigan}, {Pfuhl}, {Levin}, {Gillessen},
  {Genzel}  \& {Perets}}{{Madigan} et~al.}{2014}]{madigan+14}
{Madigan} A.-M.,  {Pfuhl} O.,  {Levin} Y.,  {Gillessen} S.,  {Genzel} R.,
  {Perets} H.~B.,  2014, \mn@doi [\apj] {10.1088/0004-637X/784/1/23}, \href
  {http://adsabs.harvard.edu/abs/2014ApJ...784...23M} {784, 23}

\bibitem[\protect\citeauthoryear{{McMillan}}{{McMillan}}{2016}]{mcmillan16}
{McMillan} B.~F.,  2016, preprint, \href
  {http://adsabs.harvard.edu/abs/2016arXiv160805580M} {} (\mn@eprint {arXiv}
  {1608.05580})

\bibitem[\protect\citeauthoryear{{Meyer} et~al.,}{{Meyer}
  et~al.}{2012}]{meyer+12}
{Meyer} L.,  et~al., 2012, \mn@doi [Science] {10.1126/science.1225506}, \href
  {http://adsabs.harvard.edu/abs/2012Sci...338...84M} {338, 84}

\bibitem[\protect\citeauthoryear{{Miyamoto} \& {Nagai}}{{Miyamoto} \&
  {Nagai}}{1975}]{M&N75}
{Miyamoto} M.,  {Nagai} R.,  1975, \pasj, \href
  {http://adsabs.harvard.edu/abs/1975PASJ...27..533M} {27, 533}

\bibitem[\protect\citeauthoryear{{Moore}, {Ghigna}, {Governato}, {Lake},
  {Quinn}, {Stadel}  \& {Tozzi}}{{Moore} et~al.}{1999}]{moore+99}
{Moore} B.,  {Ghigna} S.,  {Governato} F.,  {Lake} G.,  {Quinn} T.,  {Stadel}
  J.,   {Tozzi} P.,  1999, \mn@doi [\apjl] {10.1086/312287}, \href
  {http://adsabs.harvard.edu/abs/1999ApJ...524L..19M} {524, L19}

\bibitem[\protect\citeauthoryear{{Muno}, {Pfahl}, {Baganoff}, {Brandt}, {Ghez},
  {Lu}  \& {Morris}}{{Muno} et~al.}{2005}]{muno+05}
{Muno} M.~P.,  {Pfahl} E.,  {Baganoff} F.~K.,  {Brandt} W.~N.,  {Ghez} A.,
  {Lu} J.,   {Morris} M.~R.,  2005, \mn@doi [\apjl] {10.1086/429721}, \href
  {http://adsabs.harvard.edu/abs/2005ApJ...622L.113M} {622, L113}

\bibitem[\protect\citeauthoryear{{Navarro}, {Frenk}  \& {White}}{{Navarro}
  et~al.}{1996}]{nfw96}
{Navarro} J.~F.,  {Frenk} C.~S.,   {White} S.~D.~M.,  1996, \mn@doi [\apj]
  {10.1086/177173}, \href {http://adsabs.harvard.edu/abs/1996ApJ...462..563N}
  {462, 563}

\bibitem[\protect\citeauthoryear{{Ott}, {Eckart}  \& {Genzel}}{{Ott}
  et~al.}{1999}]{ott+99}
{Ott} T.,  {Eckart} A.,   {Genzel} R.,  1999, \mn@doi [\apj] {10.1086/307712},
  \href {http://adsabs.harvard.edu/abs/1999ApJ...523..248O} {523, 248}

\bibitem[\protect\citeauthoryear{{Paumard} et~al.,}{{Paumard}
  et~al.}{2006}]{paumard+06}
{Paumard} T.,  et~al., 2006, \mn@doi [\apj] {10.1086/503273}, \href
  {http://adsabs.harvard.edu/abs/2006ApJ...643.1011P} {643, 1011}

\bibitem[\protect\citeauthoryear{{Perets} \& {{\v S}ubr}}{{Perets} \& {{\v
  S}ubr}}{2012}]{perets12}
{Perets} H.~B.,  {{\v S}ubr} L.,  2012, \mn@doi [\apj]
  {10.1088/0004-637X/751/2/133}, \href
  {http://adsabs.harvard.edu/abs/2012ApJ...751..133P} {751, 133}

\bibitem[\protect\citeauthoryear{{Perets}, {Wu}, {Zhao}, {Famaey}, {Gentile}
  \& {Alexander}}{{Perets} et~al.}{2009}]{perets+09}
{Perets} H.~B.,  {Wu} X.,  {Zhao} H.~S.,  {Famaey} B.,  {Gentile} G.,
  {Alexander} T.,  2009, \mn@doi [\apj] {10.1088/0004-637X/697/2/2096}, \href
  {http://adsabs.harvard.edu/abs/2009ApJ...697.2096P} {697, 2096}

\bibitem[\protect\citeauthoryear{{Pfuhl} et~al.,}{{Pfuhl}
  et~al.}{2011}]{pfuhl+11}
{Pfuhl} O.,  et~al., 2011, \mn@doi [\apj] {10.1088/0004-637X/741/2/108}, \href
  {http://adsabs.harvard.edu/abs/2011ApJ...741..108P} {741, 108}

\bibitem[\protect\citeauthoryear{{Pfuhl}, {Alexander}, {Gillessen}, {Martins},
  {Genzel}, {Eisenhauer}, {Fritz}  \& {Ott}}{{Pfuhl} et~al.}{2014}]{pfuhl+14}
{Pfuhl} O.,  {Alexander} T.,  {Gillessen} S.,  {Martins} F.,  {Genzel} R.,
  {Eisenhauer} F.,  {Fritz} T.~K.,   {Ott} T.,  2014, \mn@doi [\apj]
  {10.1088/0004-637X/782/2/101}, \href
  {http://adsabs.harvard.edu/abs/2014ApJ...782..101P} {782, 101}

\bibitem[\protect\citeauthoryear{{Portail}, {Wegg}, {Gerhard}  \&
  {Martinez-Valpuesta}}{{Portail} et~al.}{2015}]{portail+15}
{Portail} M.,  {Wegg} C.,  {Gerhard} O.,   {Martinez-Valpuesta} I.,  2015,
  \mn@doi [\mnras] {10.1093/mnras/stv058}, \href
  {http://adsabs.harvard.edu/abs/2015MNRAS.448..713P} {448, 713}

\bibitem[\protect\citeauthoryear{{Price-Whelan}, {Hogg}, {Johnston}  \&
  {Hendel}}{{Price-Whelan} et~al.}{2014}]{Price2014}
{Price-Whelan} A.~M.,  {Hogg} D.~W.,  {Johnston} K.~V.,   {Hendel} D.,  2014,
  \mn@doi [\apj] {10.1088/0004-637X/794/1/4}, \href
  {http://adsabs.harvard.edu/abs/2014ApJ...794....4P} {794, 4}

\bibitem[\protect\citeauthoryear{{Rimoldi}, {Portegies Zwart}  \&
  {Rossi}}{{Rimoldi} et~al.}{2016}]{rimoldi+16}
{Rimoldi} A.,  {Portegies Zwart} S.,   {Rossi} E.~M.,  2016, \mn@doi
  [Computational Astrophysics and Cosmology] {10.1186/s40668-016-0015-4}, \href
  {http://adsabs.harvard.edu/abs/2016ComAC...3....2R} {3, 2}

\bibitem[\protect\citeauthoryear{{Rix} \& {Bovy}}{{Rix} \&
  {Bovy}}{2013}]{rix&bovy13}
{Rix} H.-W.,  {Bovy} J.,  2013, \mn@doi [\aapr] {10.1007/s00159-013-0061-8},
  \href {http://adsabs.harvard.edu/abs/2013A%26ARv..21...61R} {21, 61}

\bibitem[\protect\citeauthoryear{{Rossi}, {Kobayashi}  \& {Sari}}{{Rossi}
  et~al.}{2014}]{rossi+14}
{Rossi} E.~M.,  {Kobayashi} S.,   {Sari} R.,  2014, \mn@doi [\apj]
  {10.1088/0004-637X/795/2/125}, \href
  {http://adsabs.harvard.edu/abs/2014ApJ...795..125R} {795, 125}

\bibitem[\protect\citeauthoryear{{Sana} et~al.,}{{Sana} et~al.}{2012}]{sana+12}
{Sana} H.,  et~al., 2012, \mn@doi [Science] {10.1126/science.1223344}, \href
  {http://adsabs.harvard.edu/abs/2012Sci...337..444S} {337, 444}

\bibitem[\protect\citeauthoryear{{Sana} et~al.,}{{Sana} et~al.}{2013}]{sana+13}
{Sana} H.,  et~al., 2013, \mn@doi [\aap] {10.1051/0004-6361/201219621}, \href
  {http://adsabs.harvard.edu/abs/2013A%26A...550A.107S} {550, A107}

\bibitem[\protect\citeauthoryear{{Sari}, {Kobayashi}  \& {Rossi}}{{Sari}
  et~al.}{2010}]{skr10}
{Sari} R.,  {Kobayashi} S.,   {Rossi} E.~M.,  2010, \mn@doi [\apj]
  {10.1088/0004-637X/708/1/605}, \href
  {http://adsabs.harvard.edu/abs/2010ApJ...708..605S} {708, 605}

\bibitem[\protect\citeauthoryear{{Schaller} et~al.,}{{Schaller}
  et~al.}{2015}]{schaller+15}
{Schaller} M.,  et~al., 2015, \mn@doi [\mnras] {10.1093/mnras/stv1067}, \href
  {http://adsabs.harvard.edu/abs/2015MNRAS.451.1247S} {451, 1247}

\bibitem[\protect\citeauthoryear{{Schaye} et~al.,}{{Schaye}
  et~al.}{2015}]{schaye+15}
{Schaye} J.,  et~al., 2015, \mn@doi [\mnras] {10.1093/mnras/stu2058}, \href
  {http://adsabs.harvard.edu/abs/2015MNRAS.446..521S} {446, 521}

\bibitem[\protect\citeauthoryear{{Sch{\"o}del}, {Feldmeier}, {Neumayer},
  {Meyer}  \& {Yelda}}{{Sch{\"o}del} et~al.}{2014a}]{schodel+14rev}
{Sch{\"o}del} R.,  {Feldmeier} A.,  {Neumayer} N.,  {Meyer} L.,   {Yelda} S.,
  2014a, \mn@doi [Classical and Quantum Gravity]
  {10.1088/0264-9381/31/24/244007}, \href
  {http://adsabs.harvard.edu/abs/2014CQGra..31x4007S} {31, 244007}

\bibitem[\protect\citeauthoryear{{Sch{\"o}del}, {Feldmeier}, {Kunneriath},
  {Stolovy}, {Neumayer}, {Amaro-Seoane}  \& {Nishiyama}}{{Sch{\"o}del}
  et~al.}{2014b}]{schodel+14}
{Sch{\"o}del} R.,  {Feldmeier} A.,  {Kunneriath} D.,  {Stolovy} S.,  {Neumayer}
  N.,  {Amaro-Seoane} P.,   {Nishiyama} S.,  2014b, \mn@doi [\aap]
  {10.1051/0004-6361/201423481}, \href
  {http://adsabs.harvard.edu/abs/2014A%26A...566A..47S} {566, A47}

\bibitem[\protect\citeauthoryear{{Sesana}, {Haardt}  \& {Madau}}{{Sesana}
  et~al.}{2007}]{sesana+07}
{Sesana} A.,  {Haardt} F.,   {Madau} P.,  2007, \mn@doi [\mnras]
  {10.1111/j.1745-3933.2007.00331.x}, \href
  {http://adsabs.harvard.edu/abs/2007MNRAS.379L..45S} {379, L45}

\bibitem[\protect\citeauthoryear{{Tauris}}{{Tauris}}{2015}]{tauris15}
{Tauris} T.~M.,  2015, \mn@doi [\mnras] {10.1093/mnrasl/slu189}, \href
  {http://adsabs.harvard.edu/abs/2015MNRAS.448L...6T} {448, L6}

\bibitem[\protect\citeauthoryear{{Vera-Ciro} \& {Helmi}}{{Vera-Ciro} \&
  {Helmi}}{2013}]{V&H13}
{Vera-Ciro} C.,  {Helmi} A.,  2013, \mn@doi [\apjl]
  {10.1088/2041-8205/773/1/L4}, \href
  {http://adsabs.harvard.edu/abs/2013ApJ...773L...4V} {773, L4}

\bibitem[\protect\citeauthoryear{{Vera-Ciro}, {Helmi}, {Starkenburg}  \&
  {Breddels}}{{Vera-Ciro} et~al.}{2013}]{vera+13}
{Vera-Ciro} C.~A.,  {Helmi} A.,  {Starkenburg} E.,   {Breddels} M.~A.,  2013,
  \mn@doi [\mnras] {10.1093/mnras/sts148}, \href
  {http://adsabs.harvard.edu/abs/2013MNRAS.428.1696V} {428, 1696}

\bibitem[\protect\citeauthoryear{{Wang}, {Han}, {Cooper}, {Cole}, {Frenk}  \&
  {Lowing}}{{Wang} et~al.}{2015}]{wang+15}
{Wang} W.,  {Han} J.,  {Cooper} A.~P.,  {Cole} S.,  {Frenk} C.,   {Lowing} B.,
  2015, \mn@doi [\mnras] {10.1093/mnras/stv1647}, \href
  {http://adsabs.harvard.edu/abs/2015MNRAS.453..377W} {453, 377}

\bibitem[\protect\citeauthoryear{{Williams} \& {Evans}}{{Williams} \&
  {Evans}}{2015}]{W&E15}
{Williams} A.~A.,  {Evans} N.~W.,  2015, \mn@doi [\mnras]
  {10.1093/mnras/stv1967}, \href
  {http://adsabs.harvard.edu/abs/2015MNRAS.454..698W} {454, 698}

\bibitem[\protect\citeauthoryear{{Yu} \& {Madau}}{{Yu} \&
  {Madau}}{2007}]{yu&madau07}
{Yu} Q.,  {Madau} P.,  2007, \mn@doi [\mnras]
  {10.1111/j.1365-2966.2007.12034.x}, \href
  {http://adsabs.harvard.edu/abs/2007MNRAS.379.1293Y} {379, 1293}

\bibitem[\protect\citeauthoryear{{Zhang}, {Lu}  \& {Yu}}{{Zhang}
  et~al.}{2013}]{zhang+13}
{Zhang} F.,  {Lu} Y.,   {Yu} Q.,  2013, \mn@doi [\apj]
  {10.1088/0004-637X/768/2/153}, \href
  {http://adsabs.harvard.edu/abs/2013ApJ...768..153Z} {768, 153}

\bibitem[\protect\citeauthoryear{{Zheng} et~al.,}{{Zheng}
  et~al.}{2014}]{zheng+14}
{Zheng} Z.,  et~al., 2014, \mn@doi [\apjl] {10.1088/2041-8205/785/2/L23}, \href
  {http://adsabs.harvard.edu/abs/2014ApJ...785L..23Z} {785, L23}

\makeatother
\end{thebibliography}


\appendix

\section{Markov Chain Monte Carlo to fit the observed circular velocity} 
\label{mcmc_appendix}

\begin{figure*}
\centering
\subfigure{%
  \includegraphics[width=9.25cm]{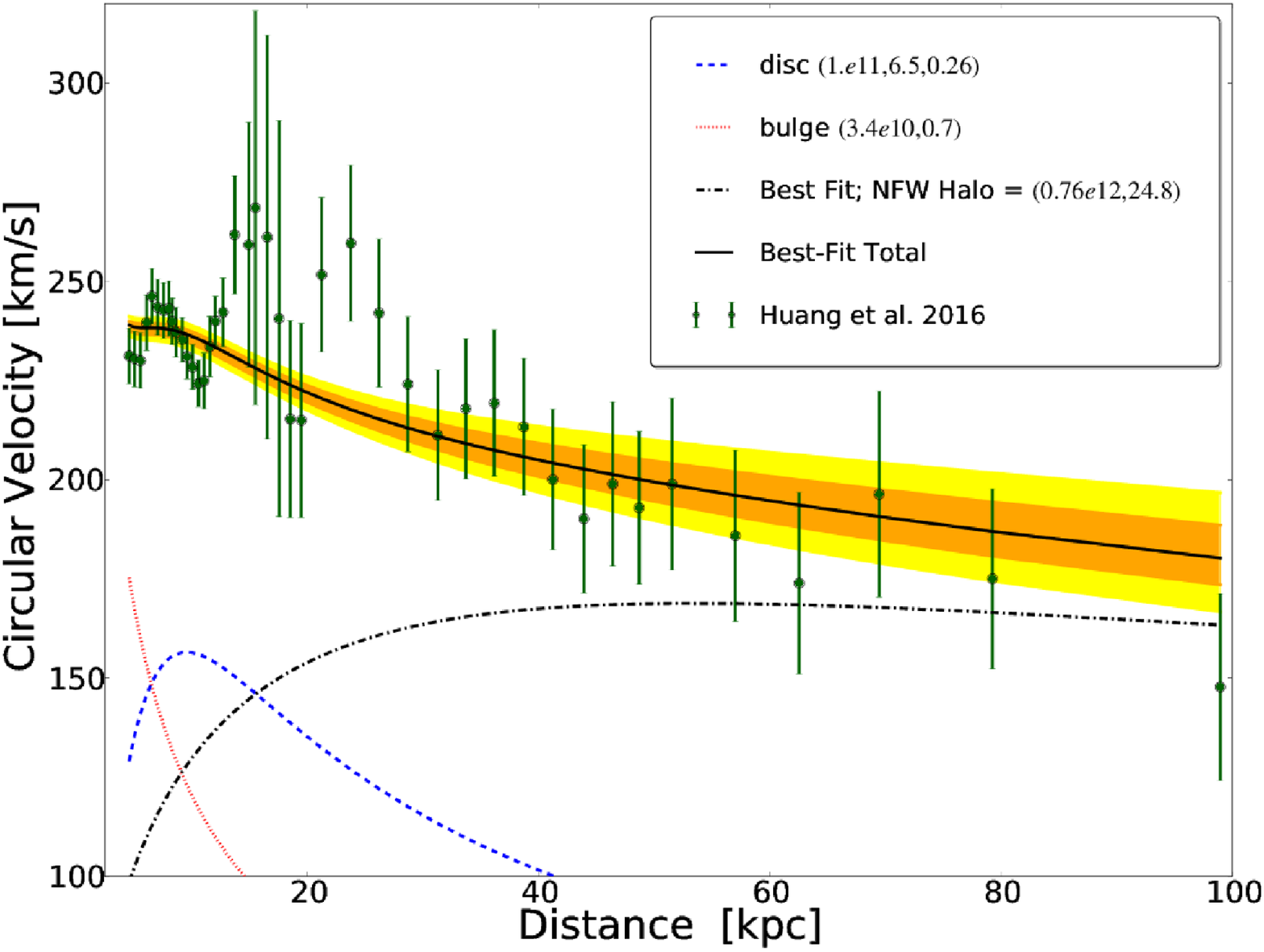}
  \label{fig:Vcirc}}
\quad
\subfigure{%
  \includegraphics[width=6.75cm]{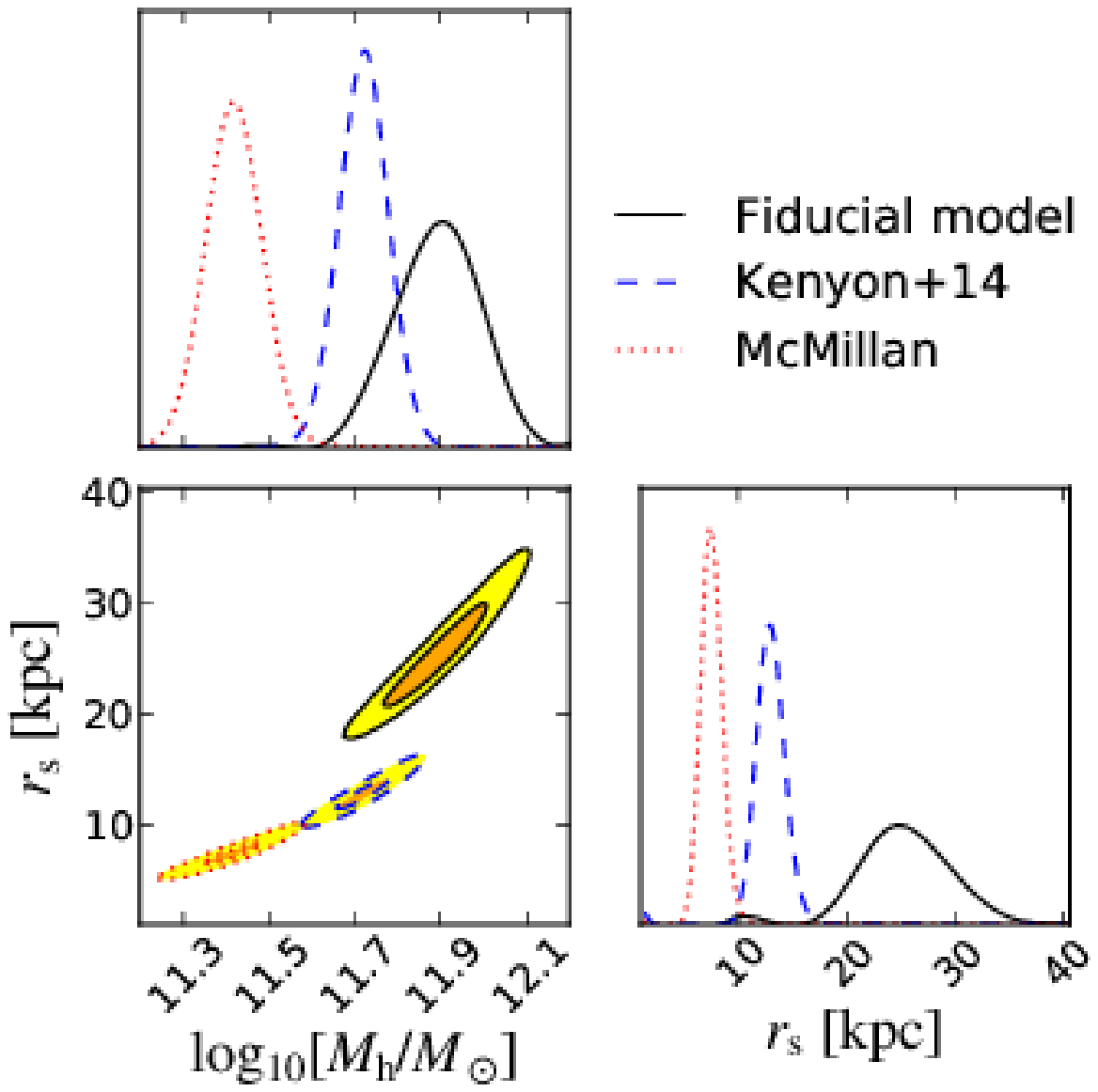}
  \label{fig:cornerplot}}
\caption{{\em Left panel:} Galactic circular velocity. Data points with error bars are
  taken from \citet{huang+16}. The orange and yellow regions
  correspond to the 68th and 95th credibility interval obtained with
  the MCMC described in the text for our fiducial Galactic Potential model. Red dotted and blue 
dashed lines
  represent the contribution from the bulge and the disc,
  respectively, whereas the dash-dotted black line indicates the
  contribution from the best-fitting NFW halo. The solid black line
  corresponds to the total circular velocity for the best-fitting model
  ($\chi^2_{\rm red} = 0.95$).  {\em Right
    panel:} Posterior distributions of the two halo parameters,
  ${\rm log_{10}[M_h/M_\odot]}$ and $r_{\rm s}$, as obtained from
  the MCMC used to fit the Galaxy circular velocity measurements
  with the three models discussed in the text (see also legend). The diagonal panels show the
  the posterior distributions for each
  parameter. 
  The lower left panel shows
  the two-dimensional marginalised posterior distributions. As
  expected, the two parameters are strongly degenerate. Orange
  (yellow) region indicates the extent of the 68\% (95\%) credibility
  interval.}
  \label{fig:appendix}
\end{figure*}

To assess which ranges of the halo mass and scale radius are compatible with
current constraints of the Milky Way halo, we employ circular velocity
measurements presented in \citet{huang+16} where the rotation curve
of the Milky Way out to $\sim$ 100 kpc has been constructed using
$\sim$ 16,000 primary red clump giants in the outer disc selected from
the LAMOST Spectroscopic Survey of the Galactic Anti-centre (LSS-GAC)
and the SDSS-III/APOGEE survey, combined with $\sim$ 5700 halo K
giants selected from the SDSS/SEGUE survey. These measurements
are reported in Figure~\ref{fig:appendix} left panel as green points with error
bars.

We remind the reader that our model for the matter density (and thus
the circular velocity) of the Milky Way consists of three components:
a bulge, a disc, and an extended (dark matter) halo. While bulge and
disc dominate the circular velocity at relatively small scales (below
about 30 kpc), larger scales are dominated by the dark matter
halo. Each of these components for all models we consider is described in detail in the main 
body
of the paper (see Sections \ref{sec:NFW} and \ref{sec:bulgedisc}).  To fit the data described above 
we fix the parameters
that refers to the bulge and the disc, whereas we consider as free
parameters those related to the dark matter halo.  We remind that dark
matter halo is assumed to have a NFW matter density profile,
completely characterised by two parameters: the total halo mass,
$\Mhs$, and the scale radius, $\rss$.

The two-dimensional parameter space $(\Mhs,\rss)$ is sampled with an
affine invariant ensemble Markov Chain Monte Carlo (MCMC) sampler
\citep{Goodman10}.  Specifically, we use the publicly available code
{\sc Emcee} \citep{Foreman13}. We run {\sc Emcee} with three separate
chains with 200 walkers and 4\,500 steps per walker. Using the
resulting \mbox{2\,700\,000} model evaluations, we estimate the
parameter uncertainties. We assess the convergence of the chains by
computing the auto-correlation time \citep[see e.g.][]{Akaret13} and
finding that our chains are about a factor of 20 times longer than it
is needed
to reach 1\% precision on the mean of each fit parameter. \\

The left panel of Figure~\ref{fig:appendix} shows the circular
velocity as a function of distance from the GC. Green points with
error bars are taken from table~3 of \citet{huang+16}, whereas orange
and yellow shaded regions correspond to the 68th and 95th credibility
intervals obtained from the MCMC procedure described above for
  our fiducial model (Section \ref{sec:NFW}). Different line styles
and colours refer to the different contributions as detailed in the
legend.  The MCMC leads to a best-fit $\chi^2$ of $39.07$ with
$N_{\rm data} = 43$ data points and $N_{\rm par} = 2$ model
parameters, thus resulting in a satisfactory reduced
$\chi^2_{\rm red} = \chi^2/(N_{\rm data}-N_{\rm par} ) = 0.95$.
  Comparable level of agreement between models\footnote{A mixed model
    that combines Kenyon at al.'s disc and McMillan's bulge gives
    results very similar to that obtained with Kenyon et al. (2014)
    disc and bulge models, so we will not discuss it further.} and
  data is obtained when adopting i) a model that combines our fiducial
  disc parameters with a lighter bulge from \citet{mcmillan16}
  ($\chi^2_{\rm red} = 1.34$) or ii) \citet{kenyon+14}'s much lighter
  disc and bulge models ($\chi^2_{\rm red} = 0.88$).

  The right panels of Figure~\ref{fig:appendix} show the posterior
  distribution of the halo parameters for the three baryonic models
  mentioned above. As expected, the two halo parameters are strongly
  degenerate but the sampling strategy has nevertheless finely sampled
  the region of high likelihood. For our fiducial baryonic model, we
  find that ${\rm log}[M_{\rm h}/M_{\odot}] = 11.89 \pm 0.18$, and
  $r_{\rm s} = 25.4 \pm 7.3$ kpc, where we quote the median and errors
  are derived from the 16th and 84th percentiles. For i) instead the
  best fitting parameters are
  ${\rm log}[M_{\rm h}/M_{\odot}] = 11.42 \pm 0.06$, and
  $r_{\rm s} = 7.5^{+1.0}_{-0.9}$ kpc, while ii) gives intermediate
  results: ${\rm log}[M_{\rm h}/M_{\odot}] = 11.72 \pm 0.06$, and
  $r_{\rm s} = 12.99^{+1.4}_{-1.3}$ kpc.

\label{lastpage}

\end{document}